\documentclass[12pt]{article}

 \usepackage{enumerate}
 \usepackage{natbib}
 \usepackage{latexsym, amsbsy, amssymb,multirow, epsfig, amsmath,verbatim, bbm, enumitem}
 \usepackage{hyperref}
 \usepackage{rotating}
 \usepackage{multirow}
 \usepackage{subcaption}
 
 \usepackage{natbib,graphicx,setspace,lscape,longtable}
 \usepackage{natbib,epsfig}
 \usepackage{amsmath,amssymb,color}
 \usepackage{algorithm}
 \usepackage{algorithmic}
 \usepackage{todonotes}
 \RequirePackage[mathlines, displaymath]{lineno}

 \allowdisplaybreaks[1]

 \def\beqr{\begin{eqnarray}}
 	\def\eeqr{\end{eqnarray}}
 \def\beqrs{\begin{eqnarray*}}
 	\def\eeqrs{\end{eqnarray*}}
 \def\bep{\begin{prop}}
 	\def\eep{\end{prop}}
 \def\be{\begin{equation}}
 	\def\ee{\end{equation}}
 \def\bea{\begin{eqnarray}}
 	\def\eea{\end{eqnarray}}
 \def\l{\left}
 \def\r{\right}
 
 \def\mR{\mathbb{R}}

 \def\indic{\mathbbm{1}} 

 \def\pr{\mbox{Pr}}
 
 \def\var{\mbox{\rm var}}
 \def\cov{\mbox{\rm cov}}

 \def\hDash{\bot\!\!\!\bot}
 \def\wh{\widehat}
 \def\wt{\widetilde}

 \def\B{{\bf B}}

 \def\u{{\bf u}}
 \def\v{{\bf v}}

 \def\w{{\bf w}}
 
 \def\x{{\bf x}}
 
 \def\y{{\bf y}}
 
 \def\z{{\bf z}}

 \newcommand{\trans}{^{\mbox{\tiny{T}}}}
 \def\defby{\stackrel{\mbox{\textrm{\rm\tiny def}}}{=}}

 \def\ep{\epsilon}
 
 \newtheorem{theorem}{Theorem}
 
  \newtheorem{proposition}{Proposition}

\begin{document}
\renewcommand{\baselinestretch}{1.3}

\title {\bf  A Distribution Free Conditional Independence Test with Applications to Causal Discovery
}
 
 \author{Zhanrui Cai$^1$, Runze Li$^2$, and Yaowu Zhang$^3$\\
           	{$^{1,2}$ The Pennsylvania State University, USA}\\
           	{$^3$ Shanghai University of Finance and Economics, China.}
           }
           
\date{\empty}

\maketitle
\renewcommand{\baselinestretch}{1.5}
\baselineskip=24pt

\begin{abstract}
	This paper is concerned with test of the conditional independence. We first establish an
	equivalence between the conditional independence and the mutual independence. Based on the equivalence, we
	propose an index to measure the conditional dependence by quantifying the mutual
	dependence among the transformed variables. The proposed index has several
	appealing properties. (a) It is distribution free since the limiting null distribution of
	the proposed index does not depend on the population distributions of the data. Hence the critical
	values can be tabulated by simulations. (b) The proposed index ranges from
	zero to one, and equals zero if and only if the conditional independence holds. Thus,
	it has nontrivial power under the alternative hypothesis. (c) It is robust to outliers and heavy-tailed data since it is invariant to conditional strictly monotone transformations. (d) It has low computational cost since it  incorporates a simple closed-form expression and can be implemented in quadratic time. (e) It is insensitive to tuning parameters involved in the calculation of the proposed index.
	(f) The new index is applicable for multivariate random vectors as well as for discrete data.
	All these properties enable us to use the new index as statistical inference tools for various data.
	The effectiveness of the method is illustrated through extensive simulations and a real application on causal discovery.

\end{abstract}
\par \vspace{9pt} \noindent {\it Key words : Conditional independence, mutual independence, distribution free.}

\pagestyle{plain}
\newpage

\section{Introduction}

Conditional independence is fundamental in graphical models and causal
inference \citep{jordan1998learning}.  Under multinormality assumption,
conditional independence is equivalent to the corresponding partial correlation
being 0. Thus, partial correlation may be used to measure conditional
dependence \citep{lawrance1976conditional}. However, partial correlation has
low power in detecting conditional dependence in the presence of nonlinear
dependence. In addition, it cannot control Type I error when the multinormality
assumption is violated. In general, testing for conditional independence is
much more challenging than for unconditional independence \citep{zhang2011kci,shah2020hardness}.

Recent works on test of conditional independence have focused on developing
omnibus conditional independence test without assuming specific functional forms of the dependencies.
\cite{linton1996conditional} proposed a nonparametric conditional independence
test based on the generalization of empirical distribution function, and
proposed using bootstrap to obtain the null distribution of the proposed test.
This diminishes the computational efficiency. Other approaches include
measuring the difference between conditional characteristic functions
\citep{su2007consistent}, the weighted Hellinger distance
\citep{su2008nonparametric}, and the empirical likelihood
\citep{su2014testing}.  Although these authors established the asymptotical
normality of the proposed test under conditional independence, the performance
of their proposed tests relies heavily on consistent estimate of the bias and
variance terms, which are quite complicated in practice. The asymptotical null
distribution may perform badly with a small sample. Thus, the authors recommended obtaining critical values of
the proposed tests by a bootstrap. This results in heavy
computation burdens. \cite{huang2010testing} proposed a test of conditional
independence based on the maximum nonlinear conditional correlation. By
discretizing the conditioning set into a set of bins, the author transforms the
original problem into an unconditional testing problem. \cite{zhang2011kci}
proposed a kernel-based conditional independence test, which essentially tests
for zero Hilbert-Schmidt norm of the partial cross-covariance operator in the
reproducing kernel Hilbert spaces. The test also required a bootstrap to
approximate the null distribution. \cite{wang2015conditional} introduced the
energy statistics into the conditional test and developed the
conditional distance correlation based on \cite{szekely2007measuring}, which can also be linked to kernel-based approaches. But the test statistics requires to compute high order U--statistics
and therefore suffers heavy computation burden, which is of order $O(n^3)$ for
a sample with size $n$. \cite{runge2018conditional} proposed a
non-parametric conditional independence testing based on the information theory
framework, in which the conditional mutual information was estimated directly
via combining the $k$-nearest neighbor estimator with a nearest-neighbor local
permutation scheme. However, the theoretical distribution of the proposed test
is unclear.

In this paper, we develop a new methodology to test conditional independence
and propose  conditional independence tests that are applicable for
continuous or discrete random variables or vectors.
Let $X$, $Y$ and $Z$ be three
continuous random variables. We are interested in testing whether $X$ and $Y$
are statistically independent given $Z$:
\beqrs
H_{0} :X\hDash Y\mid Z, \quad \textrm{ versus } \quad  H_{1} : \textrm{ otherwise}.
\eeqrs
Here we focus on random variables for simplicity. We will consider test of conditional
independence for random vectors in Section 3. To begin with, we observe that with
Rosenblatt transformation \citep{rosenblatt1952remarks}, i.e., $U\defby
F_{X\mid Z}(X\mid Z)$, $V\defby F_{Y\mid Z}(Y\mid Z)$ and $W\defby F_Z(Z)$,
$X\hDash Y\mid Z$ is equivalent to the mutual independence of $U$, $V$ and $W$.
Thus we convert a conditional independence test into a mutual independence
test, and any technique for testing mutual independence can be readily applied.
For example, \cite{chakraborty2019distance} proposed the joint distance
covariance to test mutual independence and \cite{drton2018high} constructed a
family of tests with maxima of rank correlations in high dimensions. However,
these mutual independence tests do not consider the intrinsic properties of
$U$, $V$ and $W$. This motivates us to develop a new index $\rho$ to measure
the mutual dependence.  We show that the index $\rho$ has a closed form, which
is much simpler than that of \cite{chakraborty2019distance}. In addition, it is
symmetric, invariant to strictly monotone transformations, and ranges from zero
to one, and is equal to zero if and only if $U$, $V$ and $W$ are mutually
independent. Based on the index $\rho$, we further proposed tests of
conditional independence.  We would like to further note a recent work proposed by \cite{zhou2020test}, who suggested to simply test whether $U$ and $V$ are independent. However, this is not fully equivalent to the conditional independence test and it is unclear what kind of power loss
one might have.

The proposed tests have several appealing features.
(a) The proposed test is distribution free in the sense that
its limiting null distribution does not depend on unknown parameters and
the population distributions of the data.
The fact that both $U$ and
$V$ are independent of $W$ makes the test statistic $n$-consistent under the null
hypothesis without requiring under-smoothing.  In addition, even though the test statistic depends on $U$, $V$ and $W$,
which needs to be estimated nonparametrically, we show that the test statistic has the same asymptotic properties as the
statistics where true $U$, $V$ and $W$ are directly available.
This leads to a distribution free test statistic when further considering that
$U,V$ and $W$ are uniformly distributed. Although some tests in the literature
are also distribution free, the asymptotic distributions are either complicated
to estimate \cite[e.g.,][]{su2007consistent} or rely on the Gaussian process,
which is not known how to simulate \cite[e.g.,][]{song2009testing} and would require a wild bootstrap method to determine the critical values.
Compared with existing ones, the limiting null distribution of the proposed
test depends on $U$, $V$ and $W$ only, and the critical values can be easily obtained
by a simulation-based procedure.
(b) The proposed test has nontrivial power against all fixed alternatives.
The population version of the  test statistic ranges from zero to one and equals zero if and only if conditional independence holds. Unlike many
testing procedures that are weaker than that for conditional independence
\cite[e.g.,][]{song2009testing}, the equivalence between conditional
independence and mutual independence guarantees that the newly proposed test has
nontrivial power against all fixed alternatives.
(c) The proposed test is robust since it is invariant to strictly monotone
transformations and thus, it is robust to outliers. Furthermore,
$U$, $V$ and $W$ all have bounded support, and therefore
it is suitable for handling heavy-tailed data.
(d) The proposed test has low computational cost. It is a
$V$--statistic, and direct calculation requires only $O(n^2)$ computational
complexity.
(e) It is insensitive to tuning parameters involved in the test statistics.
The test statistics are $n$--consistent under the null hypothesis without
under-smoothing, and is hence much less sensitive to the bandwidth.
The proposed index $\rho$ is extended to continuous random vectors and discrete
data in Section 3. All these properties enable us to use the new conditional independence
test for various data.

The rest of this paper is organized as follows. In Section 2 we  first show the
equivalence between conditional independence and mutual independence. We
propose a new index to measure the mutual dependence, and derive desirable
properties of the proposed index in Section 2.1. We propose an estimator for
the new index in Section 2.2. The asymptotic distributions of the proposed
estimator under the null hypothesis, global alternative, and local alternative
hypothesis are derived in Section 2.2. We extend the new index to the
multivariate and discrete cases in Section 3. We conduct numerical comparisons
and apply the proposed test to causal discovery in directed acyclic graphs in
Section 4. Some final remarks are given in Section 5. We provide some additional simulation results
as well as all the technical proofs in the appendix.

\section{Methodology}
To begin with, we establish an equivalence between conditional
independence and mutual independence. In this section, we focus on
the setting in which  $X$, $Y$ and $Z$ are continuous univariate
random variables, and the problem of interest is to test $X\hDash
Y\mid Z $. Throughout this section, denote
$U=F_{X\mid Z}(X\mid Z)$, $V=F_{Y\mid Z}(Y\mid Z)$ and $W= F_Z(Z)$.
The proposed methodology is built upon the following proposition.
\begin{proposition}\label{equivalence}
	{\color{black}  Suppose that  $X$ and $Y$ are both univariate and have continuous conditional distribution functions for every given value of $Z$, and $Z$ is a continuous univariate random variable. Then $X\hDash Y\mid Z$ if and only if $U,$ $V$ and $W$ are mutually independent.}
\end{proposition}
We provide a detailed proof of Proposition \ref{equivalence} in the appendix. Essentially, it establishes an equivalence between the conditional independence of $X\hDash Y\mid Z $ and the mutual independence among $U$, $V$ and $W$ under
the conditions in Proposition \ref{equivalence}.
Therefore, we can alleviate the hardness issue of conditional independence testing \citep{shah2020hardness}
by restricting the distribution family of the data such that $U$, $V$ and $W$ can be estimated sufficiently well using samples. As shown in our theoretical analysis, we further impose certain smoothness conditions on the conditional distributions of $X, Y\mid Z=z$ as $z$ varies in the support of $Z$. This distribution family is also considered in \cite{neykov2020minimax} to develop a minimax optimal conditional independence test.
We discuss the extension of Proposition \ref{equivalence} to  multivariate and discrete data  in Section 3.

According to Proposition \ref{equivalence}, any techniques for testing mutual
independence among three random variables can be readily applied for conditional independence testing problems. For example,  \cite{chakraborty2019distance} proposed the joint distance covariance and \cite{patra2016nonparametric} developed a bootstrap procedure to test mutual independence with known marginals.
However, a direct application of these metrics
may not be a good choice because it ignores the fact that the variables $U$, $V$ and $W$ are all uniformly distributed, as well as $U\hDash W$ and $V\hDash W$. Next, we discuss how to develop a new mutual independence test while considering these intrinsic properties of $(U, V, W)$.

\subsection{A mutual independence test}

In this section,  we propose to characterize the conditional dependence
of $X$ and $Y$ given $Z$ through quantifying the mutual dependence among
$U$, $V$ and $W$. Although our proposed test is based on the distance
between characteristics functions, our proposed test is much simpler and has
different asymptotic distribution as well as different convergence rate from
the conditional distance correlation proposed by \cite{wang2015conditional}.
Let $\omega(\cdot)$ be an arbitrary positive weight function and $\varphi_{U,V,W}(\cdot)$, $\varphi_{U}(\cdot)$, $\varphi_{V}(\cdot)$, and $\varphi_{W}(\cdot)$ be the characteristic functions of $(U,V,W)$, $U$, $V$ and $W$, respectively. Then
\beqrs
&& \textrm{$U$, $V$ and $W$ are mutually independent}\\
&\Longleftrightarrow&\varphi_{U,V,W}(t_1,t_2,t_3) = \varphi_{U}(t_1)\varphi_{V}(t_2)\varphi_{W}(t_3) \textrm{ for all $t_1,t_2,t_3\in \mR$}\\
&\Longleftrightarrow&  \iiint\big\|\varphi_{U,V,W}(t_1,t_2,t_3) -
\varphi_{U}(t_1)\varphi_{V}(t_2)\varphi_{W}(t_3)\big\|^2\omega(t_1,t_2,t_3)dt_1
dt_2 dt_3=0, \eeqrs
where $\|\psi\|^2 = \psi\trans \overline\psi$ for a complex-valued function $\psi$ and $\overline\psi$ is the conjugate of $\psi$.
By choosing $\omega(t_1,t_2,t_3)$ to be the joint probability density function
of three independent and identically distributed standard Cauchy random variables,
the integration in the above equation has a closed form,
\beqr\nonumber
&&Ee^{-|U_1-U_2|-|V_1-V_2|-|W_1-W_2|}  - 2Ee^{-|U_1-U_3|-|V_1-V_4|-|W_1-W_2|}
\\
\label{rho_index_0}
&&\hspace{3cm}+ Ee^{- |U_1-U_2|} Ee^{-|V_1-V_2|}Ee^{-|W_1-W_2|},
\eeqr
where $(U_k,V_k,W_k)$, $k=1,\ldots,4$,  are four independent copies of $(U,V,W)$. {\color{black} Here the choice of the weight function $\omega(t_1,t_2,t_3)$ is mainly for the convenient analytic form of the integration. Different from the distance correlation \citep{szekely2007measuring}, our integration exists without any moment conditions on the data, which is more widely applicable.} Furthermore, with the fact that $U\hDash W$ and $V\hDash W$, (\ref{rho_index_0})
boils down to
\beqr\label{SUV}
E\l\{S_U(U_1,U_2)S_V(V_1,V_2)e^{-|W_1-W_2|}\r\},
\eeqr
where $S_U(U_1,U_2)$ and $S_V(V_1,V_2)$ are defined as
\beqrs
S_U(U_1,U_2) &=& E\l\{e^{-|U_1-U_2|}+e^{-|U_3-U_4|}- e^{-|U_1-U_3|}-e^{-|U_2-U_3|}\mid (U_1,U_2)\r\},\\
S_V(V_1,V_2) &=& E\l\{e^{-|V_1-V_2|}+e^{-|V_3-V_4|}- e^{-|V_1-V_3|}-e^{-|V_2-V_3|}\mid (V_1,V_2)\r\}.
\eeqrs
Recall that $U$, $V$ and $W$ are uniformly distributed on $(0,1)$. {\color{black}With further calculations based on (\ref{SUV}), we obtain a normalized index and define it as $\rho$ to measure the mutual dependence:
	\beqr\nonumber
	\rho(X,Y\mid Z)&=& c_0 E\big\{\l( e^{-|U_1-U_2|} +e^{-U_1}+e^{U_1-1}+e^{-U_2}+e^{U_2-1} + 2e^{-1}-4 \r)\\\label{rho_index}
	&& \hspace{-1.3cm}\l(e^{-|V_1-V_2|} +e^{-V_1}+e^{V_1-1}+e^{-V_2}+e^{V_2-1} + 2e^{-1}-4 \r) e^{-|W_1-W_2|}\big\},
	\eeqr
	where $c_0=(13e^{-3}-40e^{-2}+13e^{-1}) ^{-1}$. }Several appealing properties of the proposed index $\rho(X,Y\mid Z)$ are summarized in Theorem  \ref{rho_property}.

\begin{theorem}\label{rho_property}
	Suppose that the conditions in Proposition \ref{equivalence} are fulfilled.
	The index $\rho(X,Y\mid Z)$ defined in (\ref{rho_index}) has the following properties:
	
	{\color{black}  (1) $0\leq \rho(X,Y\mid Z) \leq 1$, $\rho(X,Y\mid Z)=0$ holds if
		and only if $X\hDash Y\mid Z$. Furthermore,
		if $F_{X|Z}(X|Z)=F_{Y|Z}(Y|Z)$ or $F_{X|Z}(X|Z)+F_{Y|Z}(Y|Z)=1$, then
		$\rho (X,Y\mid Z)= 1$.}
	
	(2) The index $\rho$ is symmetric conditioning on $Z$. That is,
	$\rho(X,Y\mid Z)= \rho(Y,X\mid Z)$.
	
	(3) For any strictly monotone transformations $m_1(\cdot)$, $m_2(\cdot)$ and $m_3(\cdot)$, $ \rho(X,Y\mid Z)=\rho\l\{m_1(X),m_2(Y)\mid m_3(Z)\r\}$.
\end{theorem}
The step-by-step derivation of $\rho(X,Y\mid Z)$ and proof of Theorem \ref{rho_property} are presented in the appendix. Property (1) indicates that the index $\rho$ ranges from zero to one, equals zero when the conditional independence holds, and is equal to one if $Y$ is a strictly monotone transformation of $X$ conditional on $Z$. Property (2) shows that the index $\rho$ is a symmetric measure of conditional dependence. Property (3) illustrates that the index $\rho$ is invariant to any strictly monotone transformation. In fact, $\rho$ is not only invariant to marginal strictly monotone transformations, but also invariant to strictly monotone transformations conditional on $Z$. For example, it can be verified that $ \rho(X,Y\mid Z) =  \rho[ m_1\{X-E(X\mid Z)\},Y\mid Z]$.

\subsection{Asymptotic properties}

In this section, we establish the asymptotic properties of the sample
version of the proposed index under the null and alternative hypothesis.
Consider independent and identically distributed samples
$\{X_i, Y_i, Z_i\}$, $i=1,\dots, n$. To estimate the proposed index
$\rho(X,Y\mid Z)$, we apply kernel estimator for the conditional
cumulative distribution function. Specifically, define
\begin{eqnarray*}
	&&\wh f_Z(z)=n^{-1}\sum_{i=1}^{n}K_h(z-Z_i),\\
	&&\wh U = \wh F_{X\mid Z}(x\mid z)= n^{-1}\sum_{i=1}^{n} K_h(z-Z_i)
	\indic(X_i\leq x)/\wh f_Z(z), \\
	&&\wh V = \wh F_{Y\mid Z}(y\mid z)= n^{-1}\sum_{i=1}^{n} K_h(z-Z_i)
	\indic(Y_i\leq y)/\wh f_Z(z),
\end{eqnarray*}
where $K_{h}(\cdot)=K(\cdot/h)/h$, $K(\cdot)$ is a kernel function, and $h$ is the bandwidth. Besides, we use empirical distribution function to estimate the cumulative distribution function, i.e., $\wh W = \wh F_Z(z) =n^{-1}\sum_{i=1}^{n} \indic(Z_i\leq z)$. The sample version of the index, denoted by $\wh{\rho}(X,Y\mid Z)$, is thus given by
\beqrs
\wh{\rho}(X,Y\mid Z)
&=& c_0n^{-2}\sum_{i,j} \l\{\l( e^{-|\wh U_i-\wh U_j|} +e^{-\wh U_i}+e^{\wh U_i-1}+e^{-\wh U_j}+e^{\wh U_j-1} + 2e^{-1}-4 \r)\r.\\
&&\hspace{-1cm} \l.\l(e^{-|\wh V_i-\wh V_j|} +e^{-\wh V_i}+e^{\wh V_i-1}+e^{-\wh V_j}+e^{\wh V_j-1} + 2e^{-1}-4 \r) e^{-|\wh W_i-\wh W_j|}\r\}.
\eeqrs
One can also obtain a normalized index $\rho_0$, which is a direct normalization based on (\ref{rho_index_0}) without considering $U\hDash W$ and $V\hDash W$:
\beqrs
{\rho}_0(X,Y\mid Z)&=& c_0 \l\{ E e^{-|U_1-U_2|-|V_1-V_2|-|W_1-W_2|} + 8e^{-3} \r.\\
&& \hspace{-1.8cm}\l.- 2E \l(2-e^{-U_1}-e^{U_1-1}\r)\l(2-e^{-V_1}-e^{V_1-1}\r)\l(2-e^{-W_1}-e^{W_1-1}\r) \r\}.
\eeqrs
The corresponding moment estimator is
\beqrs
\wh{\rho}_0(X,Y\mid Z)&=& c_0 \l\{n^{-2}\sum_{i,j} e^{-|\wh U_i-\wh U_j|-|\wh V_i-\wh V_j|-|\wh W_i-\wh W_j|} + 8e^{-3} \r.\\
&& \hspace{-2.2cm}\l.- 2n^{-1}\sum_{i=1}^n \l(2-e^{-\wh U_i}-e^{\wh U_i-1}\r)\l(2-e^{-\wh V_i}-e^{\wh V_i-1}\r)\l(2-e^{-\wh W_i}-e^{\wh W_i-1}\r) \r\}.
\eeqrs
Although $\rho(X,Y\mid Z) = \rho_0(X,Y\mid Z)$ at the population level, those two statistics $\wh{\rho}(X,Y\mid Z)$ and $\wh{\rho}_0(X,Y\mid Z)$ exhibit different properties at the sample level. This is because $\wh{\rho}(X,Y\mid Z)$ considers the fact that $U\hDash W$ and $V\hDash W$. But on the other hand, $\wh{\rho}_0(X,Y\mid Z)$ is only a regular mutual independence test statistic, where $\wh U_i$, $\wh V_i$, and $\wh W_i$ are exchangeable. When $X\hDash Y\mid Z$,
under Conditions 1-4 listed below, $\wh{\rho}(X,Y\mid Z)$ is of order $(n^{-1}+h^{4m})$, while $\wh{\rho}_0(X,Y\mid Z)$ is of order $(n^{-1}+h^{2m})$  because of the bias caused by nonparametric estimation. Note that $m$ is the order of kernel functions and equal to 2 when using regular kernel functions such as Gaussian and epanechnikov kernels. This indicates that $\wh{\rho}(X,Y\mid Z)$ is essentially $n$ consistent without under-smoothing while $\wh{\rho}_0(X,Y\mid Z)$ typically requires under-smoothing. In addition, our statistic $\wh{\rho}(X,Y\mid Z)$ has the same asymptotic properties as if $U,V$ and $W$ are observed, but $\wh{\rho}_0(X,Y\mid Z)$ does not. See Figure \ref{boot_distn} for a numerical comparison between the empirical null distributions of the two statistics.

We next study the asymptotical behaviors of the estimated index, $\wh{\rho}(X,Y\mid Z)$, under both the null and the alternative hypotheses. The following regularity conditions are imposed to facilitate our subsequent theoretical analyses. In what follows, we derive the limiting distribution of $\wh{\rho}(X,Y\mid Z)$ under the null hypothesis in Theorem \ref{rho_null}.

\textit{Condition}  1.
The univariate kernel function $K(\cdot)$ is symmetric about zero and Lipschitz continuous. In addition,  it satisfies
\beqrs
\int K(\upsilon)d\upsilon =1, \quad \int \upsilon^iK(\upsilon)d\upsilon =0,1\leq i \leq m-1, \quad
0\neq \int \upsilon^m K(\upsilon)d\upsilon <\infty.
\eeqrs

\textit{Condition} 2. The bandwidth  $h$  satisfies $nh^2/\log^2(n)\to \infty$, and $nh^{4m}\to0$.

\textit{Condition} 3. The probability density function of $Z$, denoted by $f_Z(z)$ is bounded away from $0$ to infinity.

\textit{Condition} 4. The $(m-1)$th derivatives of $F_{X\mid Z}(x\mid z)f(z)$, $F_{Y\mid Z}(y\mid z)f(z)$ and $f_Z(z)$ with respect to $z$ are locally Lipschitz-continuous.

{\color{black}
	\begin{theorem}\label{rho_null}
		Suppose that Conditions 1-4 hold  and the conditions in Proposition \ref{equivalence} are fulfilled. Under the null hypothesis,
		\beqrs
		n\wh{\rho}(X,Y\mid Z)\to c_0\sum_{j=1}^{\infty}\lambda_j\chi_{j}^{2}(1),
		\eeqrs
		in distribution,    where $\chi_{j}^2(1)$, $j=1, 2, \dots$ are independent chi-square
		random variables with one degree of freedom, and $\lambda_j$s, $j=1, 2, \ldots$
		are eigenvalues of
		\beqrs
		h(u,v,w;u',v',w') =&(e^{-|u-u'|}+e^{-u}+e^{u-1}+e^{-u'}+e^{u'-1} + 2e^{-1}-4)
		(e^{-|v-v'|}\\
		&+e^{-v}+e^{v-1}+e^{-v'} +e^{v'-1} + 2e^{-1}-4)  e^{-| w- w'|}.
		\eeqrs
		That is, there exists orthonormal eigenfunction $\Phi_j(u,v,w)$ such that
		
		\beqr\nonumber
		\int_0^1\int_0^1\int_0^1 h(u,v,w;u',v',w') \Phi_j(u',v',w')du'dv'dw' = \lambda_j \Phi_j(u,v,w).
		\eeqr
		
	\end{theorem}
	The proof of Theorem \ref{rho_null} is given in the appendix.
	To understand the asymptotic distributions intuitively, we showed in the proof that $n\wh{\rho}(X,Y\mid Z)$ can be approximated by degenerate V-statistics, i.e.,
	$$n\wh{\rho}(X,Y\mid Z)=c_0n^{-1}\sum_{i,j} h(U_i,V_i,W_i;U_j,V_j,W_j)+o_p(n^{-1}),$$
	and
	$E\{h(U_i,V_i,W_i;U_j,V_j,W_j)\mid (U_i,V_i,W_i)\}=0$.
	By the spectral decomposition,
	$$
	h(u,v,w;u',v',w')= \sum_{j=1}^\infty\lambda_j\Phi_j(u,v,w)\Phi_j(u',v',w').
	$$
	Therefore, $n\wh{\rho}(X,Y\mid Z)= c_0  \sum_{j=1}^\infty\lambda_j\{n^{-1/2}\sum_{i=1}^n\Phi_j(U_i,V_i,W_i) \}^2+o_p(n^{-1}),$ which  converges in distribution to the weighted sum of independent chi squared distributions provided in Theorem \ref{rho_null} because
	$n^{-1/2}\sum_{i=1}^n\Phi_j(U_i,V_i,W_i) $ is asymptotically standard normal
	\citep{korolyuk2013theory}.
	Moreover, the $\lambda_j$s, $j=1, 2, \dots$ are real numbers associated with the distribution of
	$U$, $V$ and $W$, all of which follow  uniform distributions on $[0,1]$.
	In addition, $U$, $V$ and $W$ are mutually independent under the null hypothesis.
	This indicates that the proposed test statistic is essentially distribution free under the null hypothesis.
	Therefore, we suggest a simulation procedure to approximate the null distribution and decide the critical value. The simulation procedure can be independent of the original data and hence greatly improved the computation efficiency. In what follows, we describe the simulation-based procedure in detail to decide the critical value $c_\alpha$.
	
	\begin{enumerate}
		\item Generate $\{U_i^*, V_i^*, W_i^*\}$, $i=1,\dots, n$ independently from mutually independent standard uniform distributions;
		\item Compute the statistic $\wh{\rho^*}$ based on $\{U_i^*, V_i^*, W_i^*\}$, $i=1,\dots, n$, i.e.,
		\beqr\nonumber
		\wh{\rho^*}&=& c_0n^{-2}\sum_{i,j} \l\{\l( e^{-|U_i^*-U_j^*|} +e^{-U_i^*}+e^{U_i^*-1}+e^{-U_j^*}+e^{U_j^*-1} + 2e^{-1}-4 \r)\r.\hspace{2.3cm}\\\label{rho_hat_bootstrap}
		&& \l.\l(e^{-|V_i^*-V_j^*|} +e^{-V_i^*}+e^{V_i^*-1}+e^{-V_j^*}+e^{V_j^*-1} + 2e^{-1}-4 \r) e^{-|W_i^*-W_j^*|}\r\}.
		\eeqr
		\item Repeat Steps 1-2 for $B$ times and set $c_\alpha$ to be the upper $\alpha$ quantile of the estimated $\wh{\rho^*}$ obtained from the randomly simulated samples.
	\end{enumerate}
	Because $(U^*,V^*,W^*)$ has the same distribution as that of $(U,V,W)$ under the null hypothesis, it is straightforward that this simulation-based procedure can provide a valid approximation of the asymptotic null distribution of $\wh{\rho}(X,Y\mid Z)$ when $B$ is large. The consistency of this procedure is guaranteed by Theorem \ref{rho_bootstrap}.
	
	\begin{theorem}\label{rho_bootstrap}
		Under Conditions 1-4, it follows that
		\beqrs
		n\wh{\rho^*} \to c_0\sum_{j=1}^{\infty}\lambda_j\chi_{j}^{2}(1)
		\eeqrs
		in distribution,   where $\chi_{j}^2(1)$, $j=1, 2, \dots$ are independent $\chi^2(1)$ variables, and $\lambda_j$, $j=1, 2, \dots$ are the same as that of Theorem \ref{rho_null}.
\end{theorem}}

Next, we study the power performance of the proposed test under two kinds of
alternative hypotheses, under which the conditional independence no longer holds.
We first consider the global alternative, denoted by $H_{1g}$, we have
\beqrs
H_{1g}: \quad  X \not\!\perp\!\!\!\perp Y\mid Z.
\eeqrs
We then consider a sequence of local alternatives, denoted by $H_{1l}$,
\beqrs
H_{1l} : \quad F_{X\mid Z}(x \mid Z=z)-F_{X \mid (Y,Z)}\left\{x \mid (Y=y,Z=z)\right\}=n^{-1 / 2} \ell (x,y,z).
\eeqrs
The asymptotical properties of the test statistics $\wh{\rho}(X,Y\mid Z)$ under
the global alternative and local alternatives are given in Theorem \ref{rho_power},
whose proof is in the appendix. Theorem~\ref{rho_power}
shows that the proposed test can consistently detect any fixed alternatives as well as local alternatives at rate $O(n^{-1/2})$.
\begin{theorem}\label{rho_power}
	Suppose that Conditions 1-4 hold and the conditions in Proposition \ref{equivalence} are fulfilled.
	Under $H_{1g}$,     when $nh^{2m}\to0$,
	\beqrs
	n^{1/2} \l\{\wh{\rho}(X,Y\mid Z)-{\rho}(X,Y\mid Z)\r\} \to \mathcal{N} (0,\sigma^2_0),
	\eeqrs
	in distribution,    where $\sigma_0^2\defby 4c_0^2\var(P_{1,1}+P_{2,1}+P_{3,1}+P_{4,1})$, and $(P_{1,1}, P_{2,1}, P_{3,1}, P_{4,1})$ are  defined in (\ref{P1i})-(\ref{P4i}) in the appendix, respectively.
	
	Under $H_{1l}$,
	\beqrs
	n\wh{\rho}(X,Y\mid Z)\to \iiint\left\|\zeta(t_1,t_2,t_3)\right\|^{2} \omega(t_1,t_2,t_3) dt_1dt_2dt_3,
	\eeqrs
	in distribution,    where $\zeta(t_1,t_2,t_3)$ stands for a complex-valued Gaussian random process with mean function $E\left[it_1\ell(X,Y,Z)e^{it_1 U}\left\{e^{it_2 V}-\varphi_{V}(t_2)\right\}e^{it_3 W}\right]$ and covariance function defined in (\ref{cov}) in the appendix, and $\omega(t_1,t_2,t_3)$ is the joint probability density function of three independent and identically distributed standard Cauchy random variables.
\end{theorem}

\section{Extensions}

\subsection{Multivariate continuous data}
The methodology developed in Section 2 assumes that all the variables are univariate.
In this section, we generalize the proposed index $\rho$ to the multivariate case.
Let $\x =(X_1,\ldots,X_p)\trans\in\mR^p$,
$\y = (Y_1,\ldots, Y_q)\trans\in\mR^q$ and $\z = (Z_1,\ldots, Z_r)\trans\in\mR^r$
be continuous random vectors. More specifically, all elements of $\x$, $\y$ and $\z$ are continuous
random variables. Define $F_{X_1\mid \z}(X_1\mid \z)$ for the cumulative
distribution function of $X_1$ given $\z
$ and $F_{X_k\mid \z,X_1,\ldots,X_{k-1}}(X_k\mid \z,X_1,\ldots,X_{k-1})$ for the cumulative
distribution function of $X_k$ given $\z,X_1,\ldots,X_{k-1}
$ for  $k=2,\cdots, p$. Similar notation apply for $\y$ and $\z$.
Denote
$\wt U_1=F_{X_1\mid \z}(X_1\mid \z)$, $\wt V_1=F_{Y_1\mid \z}(Y_1\mid \z)$,
$\wt W_1 = F_{Z_1}(Z_1)$,
\begin{eqnarray*}
	&&\wt U_k = F_{X_k\mid \z,X_1,\ldots,X_{k-1}}(X_k\mid \z,X_1,\ldots,X_{k-1}),
	\ k=2,\ldots,p,
	\\
	&& \wt V_k = F_{Y_k\mid \z,Y_1,\ldots,Y_{k-1}}(Y_k\mid \z,Y_1,\ldots,Y_{k-1}),
	\ k=2,\ldots,q,
	\\
	&&\wt W_k = F_{ Z_k\mid Z_1,\ldots,Z_{k-1}}( Z_k\mid Z_1,\ldots,Z_{k-1}),\ k=2,\ldots,r.
\end{eqnarray*}
{\color{black}Further denote $\u=(\wt U_1,\ldots,\wt U_p)\trans$,
	$\v=(\wt V_1,\ldots,\wt V_q)\trans$, $\w=(\wt W_1,\ldots,\wt W_r)\trans$.
	Similar to test of conditional independence for random variables,
	we first establish an equivalence between the conditional independence
	$\x\hDash\y\mid\z$ and the  mutual independence of $\u$,$\v$ and $\w$,
	which is stated in Theorem \ref{rho_multivariate}.
	
	\begin{theorem}\label{rho_multivariate}
		Assume that all the  conditional cumulative distribution functions used in constructing $\u$,$\v$ and $\w$ are continuous for every given values, then $\x\hDash\y\mid\z$ if and only if $\u$,$\v$ and $\w$ are mutually independent.
\end{theorem}}

The proof of Theorem \ref{rho_multivariate} is illustrated in the appendix. Theorem \ref{rho_multivariate} established an equivalence between
the conditional independence $\x\hDash\y\mid\z$  and the mutual independence
among $\u$, $\v$ and $\w$. It is notable that when $p,q,r$ are relatively
large, $(\u,\v,\w)$ may be difficult to estimate because of the curse of
dimensionality. In this paper, we mainly focus on the low dimensional case. {\color{black}Next, we develop the mutual independence test among $\u$,$\v$ and $\w$. Similar as the univariate case, we set the weight function to be the joint density of $(p+q+r)$ independent and identically distributed standard Cauchy random variables. We may further derive the closed form expression of $\rho(\x, \y \mid\z)$
	\beqrs
	\rho(\x, \y \mid\z) = E\l\{ S_{\u}(\u_1,\u_2)S_{\v}(\v_1,\v_2)e^{-\|\w_1-\w_2\|_1}\r\},
	\eeqrs
	where $(\u_1, \v_1, \w_1)$ and $(\u_2, \v_2, \w_2)$ are two independent
	copies of $(\u, \v, \w)$. Moreover,
	\beqrs
	S_{\u}(\u_1,\u_2) &=& E\l\{e^{-\|\u_1-\u_2\|_1}+e^{-\|\u_3-\u_4\|_1}- e^{-\|\u_1-\u_3\|_1}-e^{-\|\u_2-\u_3\|_1}\mid (\u_1,\u_2)\r\}, \\
	S_{\v}(\v_1,\v_2) &=& E\l\{e^{-\|\v_1-\v_2\|_1}+e^{-\|\v_3-\v_4\|_1}- e^{-\|\v_1-\v_3\|_1}-e^{-\|\v_2-\v_3\|_1}\mid (\v_1,\v_2)\r\}.
	\eeqrs
	and $\|\cdot\|_1$ is the $\ell_1$ norm. Then $\rho(\x, \y \mid\z)$ is nonnegative and equals zero if and only if $\x\hDash\y\mid\z$.
	By estimating $\rho(\x, \y \mid\z)$ consistently at the sample level, the resulting test is clearly consistent. To implement the test, it is still required to study the asymptotic distributions under the conditional independence using independent and identically distributed samples
	$\{\x_i, \y_i, \z_i\}$, $i=1,\dots, n$.
	We also apply kernel estimator for the conditional
	cumulative distribution functions when estimating $\u_i,\v_i$ and $\w_i$. Specifically, we estimate $F_{A\mid B_1,\ldots, B_\ell}(a\mid b_1,\ldots,b_\ell)$ with
	\begin{eqnarray*}
		&&\wh F_{A\mid B_1,\ldots, B_\ell}(a\mid b_1,\ldots,b_\ell)= \frac{\sum_{i=1}^{n} \indic(A_{i}\leq a) \prod_{k=1}^\ell K_h(B_{ik}-b_{k}) }{\sum_{i=1}^{n} \prod_{k=1}^\ell K_h(B_{ik}-b_{k})}.
	\end{eqnarray*}
	where $(A,B_1,\ldots, B_\ell)\trans$ are  $(Z_k,Z_1,\ldots,Z_{k-1})\trans$, $k=2,\ldots,r$,
	$(X_\ell,X_1,\ldots,X_{\ell-1},\z\trans)\trans$, $\ell=1,\ldots,p$, or
	$(Y_j,Y_1,\ldots,Y_{j-1},\z\trans)\trans$, $j=1,\ldots,q$ when estimating $\w$, $\u$ and $\v$, respectively. The sample version of $\rho(\x, \y \mid\z)$ is given by
	\beqrs
	\wh \rho(\x, \y \mid\z)
	&=&  n^{-2}\sum_{i,j} E\Big[\left\{e^{-\|\wh\u_i-\wh\u_j\|_1}+e^{-\|\u-\u'\|_1}- e^{-\|\wh\u_i-\u\|_1}-e^{-\|\u-\wh\u_j\|_1}\mid (\wh\u_i,\wh\u_j)\r\}\\
	&&E\left\{e^{-\|\wh\v_i-\wh\v_j\|_1}+e^{-\|\v-\v'\|_1}- e^{-\|\wh\v_i-\v\|_1}-e^{-\|\v-\wh\v_j\|_1}\mid (\wh\v_i,\wh\v_j)\r\}
	e^{-\|\wh\w_i-\wh\w_j\|_1}\Big],
	\eeqrs
	where $(\u',\v')$ is an independent copy of $(\u,\v)$,
	and further calculations yield that $\wh \rho(\x, \y \mid\z)$ is equal to
	\beqrs
	n^{-2}\sum_{i,j} \l[\l\{ e^{-\|\wh \u_i-\wh \u_j\|_1} + (\frac{2}{e})^p- \prod_{k=1}^p(2-e^{-\wh{\wt U}_{ik}-1}-e^{-\wh{\wt U}_{ik}})-\prod_{k=1}^p(2-e^{-\wh{\wt U}_{jk}-1}-e^{-\wh{\wt U}_{jk}}) \r\}\r.\\
	\l.\l\{ e^{-\|\wh \v_i-\wh \v_j\|_1} + (\frac{2}{e})^q- \prod_{k=1}^q(2-e^{-\wh{\wt V}_{ik}-1}-e^{-\wh{\wt V}_{ik}})-\prod_{k=1}^q(2-e^{-\wh{\wt V}_{jk}-1}-e^{-\wh{\wt V}_{jk}}) \r\}e^{-\|\wh \w_i-\wh \w_j\|_1}\r].
	\eeqrs
	We next study the asymptotical behaviors of $\wh \rho(\x, \y \mid\z)$ under the null hypothesis  in Theorem \ref{T_n_null}, whose proof is given in the appendix. We begin by providing some regularity conditions for the multivariate data.

	\textit{Condition} $2'$. The bandwidth  $h$  satisfies $nh^{2(r+p-1)}/\log^2(n)\to \infty$,
	$nh^{2(r+q-1)}/\log^2(n)\to \infty$, and $nh^{4m}\to0$.
	
	\textit{Condition} $3'$. The probability density function of the random vector $(Z_1,\ldots,Z_k)\trans$, $k=1,\ldots,r$, $(\z\trans, X_1,\ldots,X_\ell)\trans$, $\ell=1,\ldots,p-1$, and $(\z\trans,Y_1,\ldots,Y_m)\trans$, $m=1,\ldots,q-1$, are all bounded away from $0$ to infinity.
	
	\textit{Condition} $4'$. The $(m-1)$th derivatives of $F_{A\mid \B}(a\mid \mathbf b)f_\B(\mathbf b)$, and $f_\B(\mathbf b)$ with respect to $\mathbf b$ are locally Lipschitz-continuous, where $(A,\B\trans)\trans$ can be any one of $(Z_k,Z_1,\ldots,Z_{k-1})\trans$, $k=2,\ldots,r$,
	$(X_\ell,X_1,\ldots,X_{\ell-1},\z\trans)\trans$, $\ell=1,\ldots,p$, or
	$(Y_j,Y_1,\ldots,Y_{j-1},\z\trans)\trans$, $j=1,\ldots,q$.
	
	\begin{theorem}\label{T_n_null}
		Suppose that Conditions 1 and $2'$-$4'$ hold  and the conditions in Theorem \ref{rho_multivariate} are fulfilled. Under the null hypothesis,
		\beqrs
		n\wh \rho(\x, \y \mid\z) \to \sum_{j=1}^{\infty}\lambda_j\chi_{j}^{2}(1),
		\eeqrs
		in distribution,    where $\chi_{j}^2(1)$, $j=1, 2, \dots$ are independent chi-square random variables with one degree of freedom, and $\lambda_j$s, $j=1, 2, \ldots$ are eigenvalues of
		\beqrs
		\wt h(\u, \v, \w; \u', \v', \w') = S_\u(\u,\u') S_\v(\v,\v')e^{-\| \w- \w'\|_1}.
		\eeqrs
		That is, there exists orthonormal eigenfunction $\Phi_j(\u,\v,\w)$ such that
		\beqrs
		\iiint_{[0,1]^{p+q+r}}\wt h(\u, \v, \w; \u', \v', \w')   \Phi_j(\u',\v',\w')d\u'd\v'd\w' = \lambda_j \Phi_j(\u,\v,\w).
		\eeqrs
	\end{theorem}
}

\subsection{Discrete data}
In this section, we discuss the setting in which $X$, $Y$ and $Z$ are univariate discrete random variables. Specifically, we apply transformations in \cite{brockwell2007universal} to obtain $U$ and $V$.
Define $F_{X\mid Z}(x\mid z) = \pr(X\leq x\mid Z=z)$, $F_{X\mid Z}(x-\mid z) = \pr(X< x\mid Z=z)$,
$F_{Y\mid Z}(y\mid z) = \pr(Y\leq y\mid Z=z)$, and $F_{Y\mid Z}(y-\mid z) = \pr(Y< y\mid Z=z)$. We further let $U_X$ and $U_Y$ be two independent and identically distributed $U(0,1)$ random variables, and apply the transformations
\beqrs
U& = &(1-U_X) F_{X\mid Z}(X-\mid Z)+U_X F_{X\mid Z}(X\mid Z),\\
V& = &(1-U_Y) F_{Y\mid Z}(Y-\mid Z)+U_Y F_{Y\mid Z}(Y\mid Z).
\eeqrs
According to \cite{brockwell2007universal}, both $U$ and $V$ are uniformly distributed on $(0,1)$. In addition, $U\hDash Z$ and $V\hDash Z$. In the following proposition, we establish the equivalence between the conditional independence and the mutual independence.
\begin{theorem}\label{rho_discrete}
	For discrete random variables $X$, $Y$ and $Z$,
	$X\hDash Y\mid Z$ if and only if $U,V$ and $Z$ are mutually independent.
\end{theorem}
The proof of Theorem \ref{rho_discrete} is presented in the appendix.  With Theorem \ref{rho_discrete}, we turn a discrete conditional independence problem into a mutual independence one. Hence similar techniques can be readily applied for the mutual independence test and we omit them to avoid verbosity.


\section{Numerical Validations}
\subsection{Conditional independence test}
In this section, we investigate the finite sample performance of the proposed
methods. To begin with, we illustrate that the null distribution of
$\wh\rho(X,Y\mid Z)$ is indeed distribution free as if $U$, $V$ and $W$ can be
observed, and is insensitive to the bandwidth of the nonparametric kernel. In
comparison, the null distribution of $\wh{\rho}_0(X,Y\mid Z)$ does not enjoy
such properties. To facilitate the analysis, let $X=Z+\varepsilon_1$ and
$Y=Z+\varepsilon_2$, where $Z$, $\varepsilon_1$ and $\varepsilon_2$ are
independent and identically distributed. We consider three scenarios where $Z$,
$\varepsilon_1$, $\varepsilon_2$ are independently drawn from normal
distribution $N(0,1)$, uniform distribution $U(0,1)$, and exponential
distribution $Exp(1)$, respectively. It is clear that $X$ and $Y$ are
conditionally independent given $Z$. The sample size $n$ is set to be 100.

\begin{figure}[h!]
	\begin{tabular}{ccc}
		\psfig{figure= 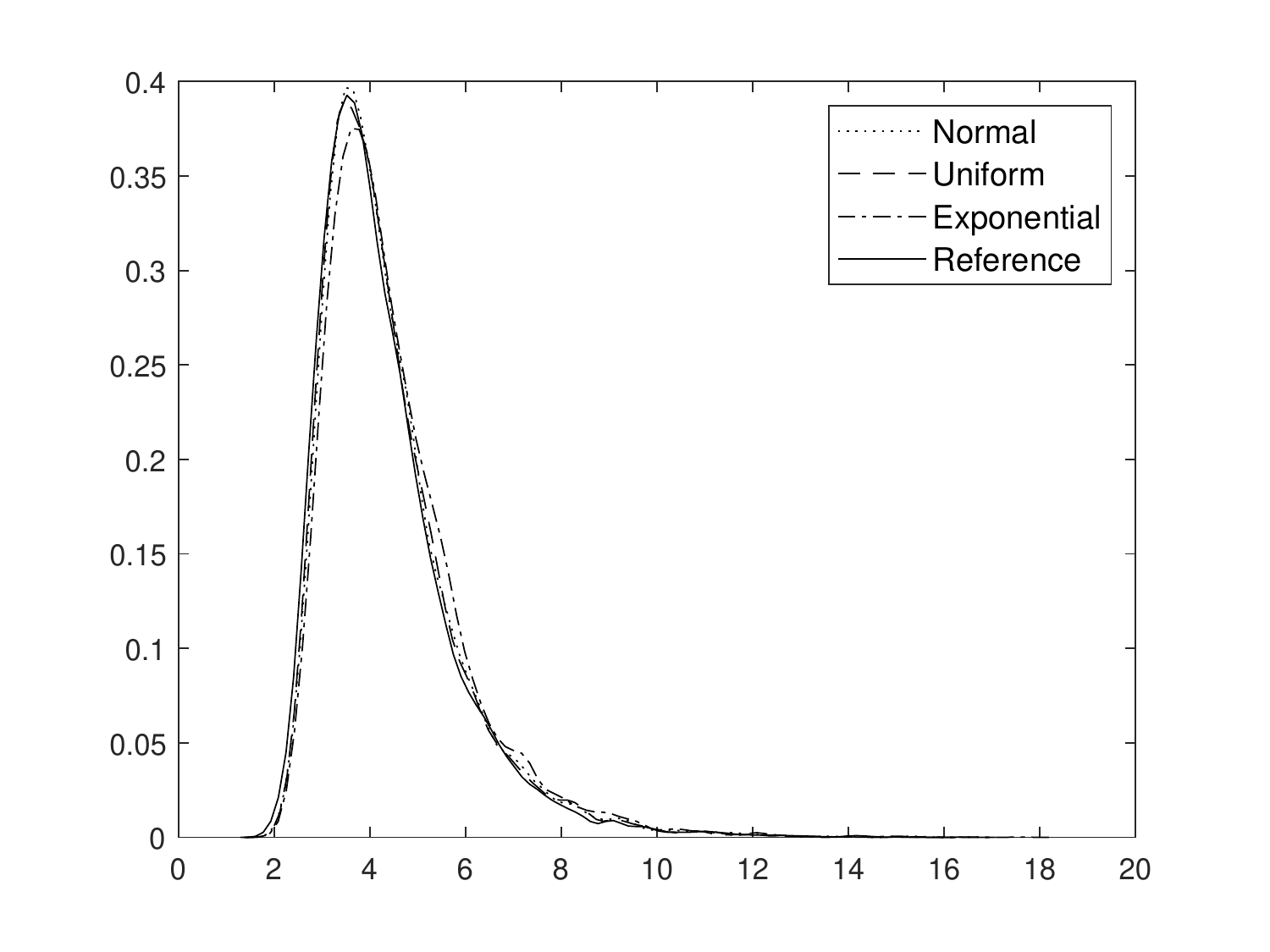,width=3in,height=2.5in,angle=0} &
		\psfig{figure= 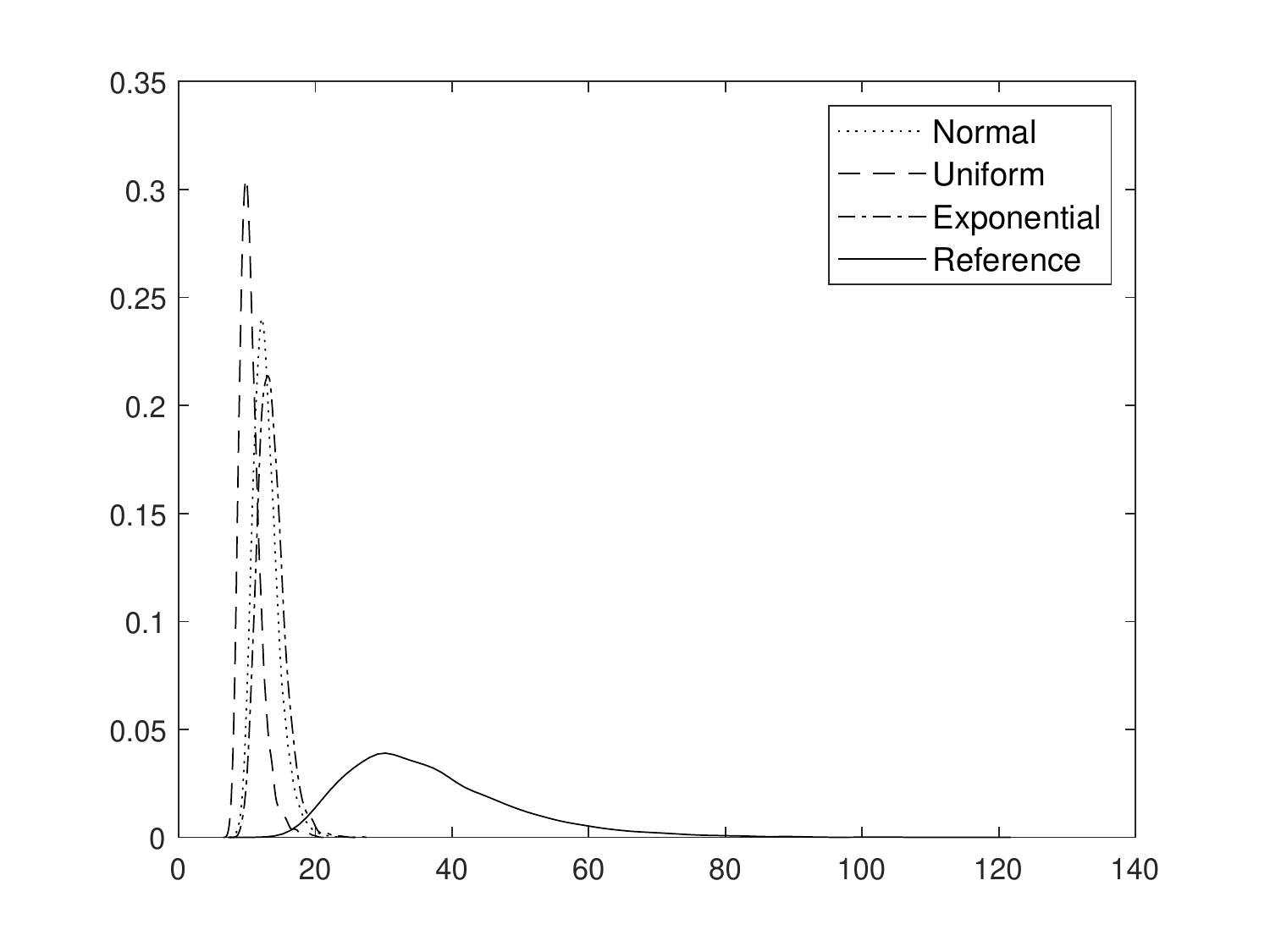,width=3in,height=2.5in,angle=0} &\\
		(a) & (b) & \\
		\psfig{figure= 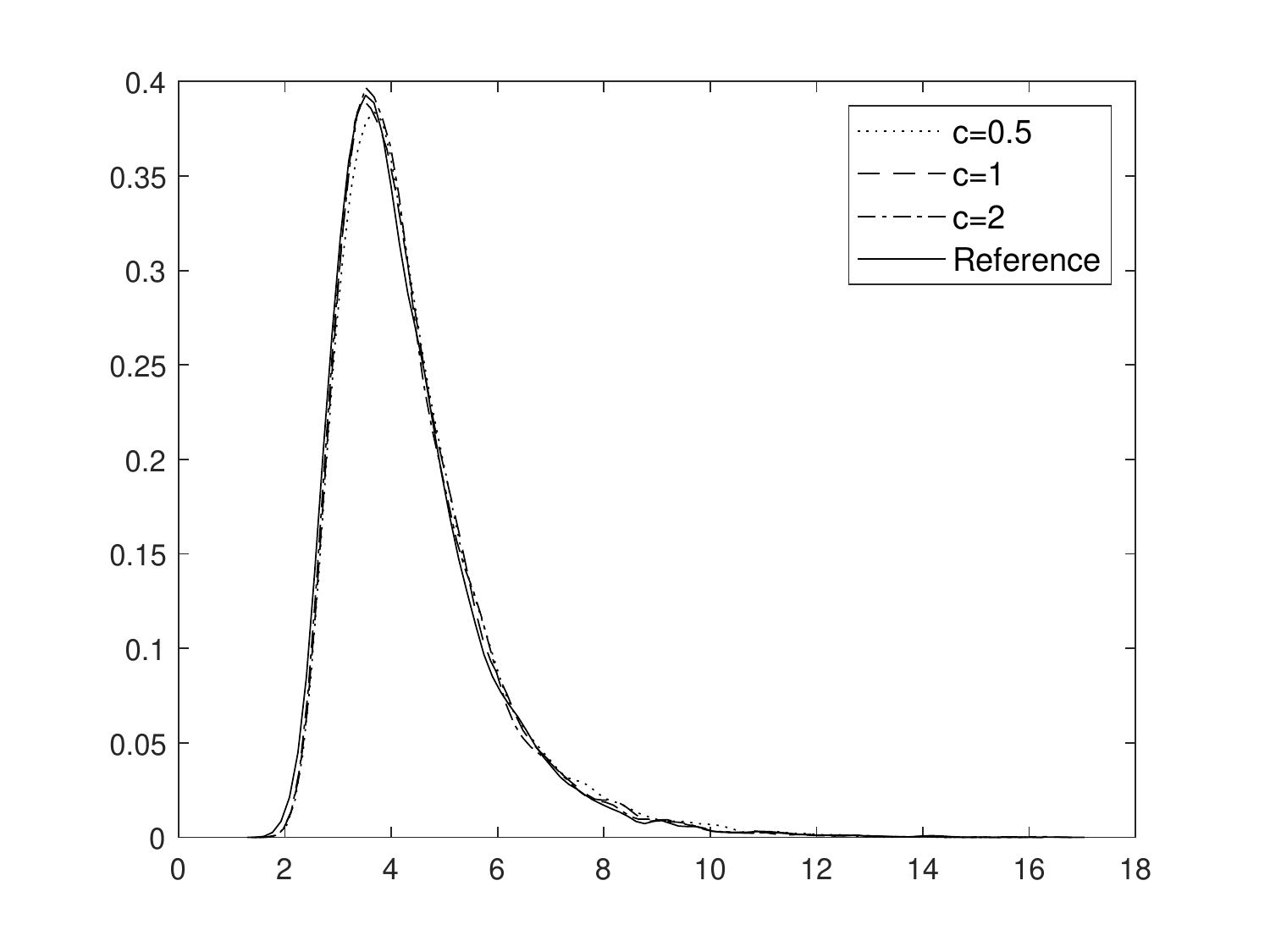,width=3in,height=2.5in,angle=0} &
		\psfig{figure= 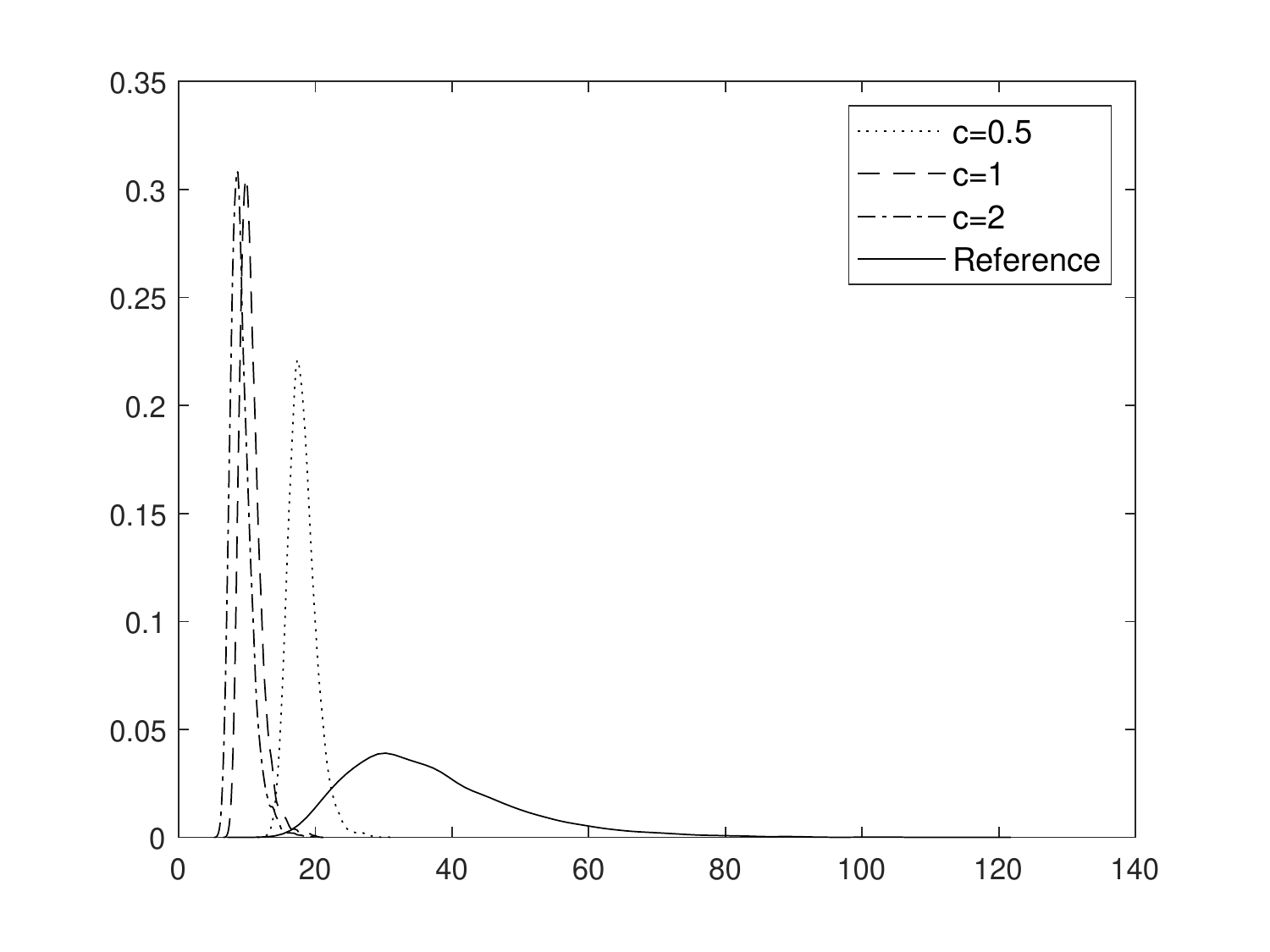,width=3in,height=2.5in,angle=0} &\\
		(c) & (d) &
	\end{tabular}
	\caption{(a) and (b) are the simulated null distributions for
		$n\wh{\rho}(X,Y\mid Z)$ and $n\wh{\rho}_0(X,Y\mid Z)$ when data were
		generated from different
		distributions, while (c) and (d) are the simulated null distributions for
		$n\wh{\rho}(X,Y\mid Z)$ and $n\wh{\rho}_0(X,Y\mid Z)$ using different bandwidths,
		respectively. }
	\label{boot_distn}
\end{figure}

The simulated null distributions based on $n\wh\rho(X,Y\mid Z)$ and
$n\wh\rho_0(X,Y\mid Z)$ are depicted in Figure~\ref{boot_distn}. The estimated kernel
density curves of $n\wh\rho(X,Y\mid Z)$ based on 1000 repetitions are shown in
Figure~\ref{boot_distn}(a), where the reference curve is generated by
the simulation-based statistic $n\wh{\rho^*}(X,Y\mid Z)$ defined in
(\ref{rho_hat_bootstrap}).
Clearly all the estimated density curves are close to the reference, indicating
that limiting null distribution of the estimated index is indeed distribution
free as if no kernel estimation is involved. In comparison, we apply the same
simulation settings for $\wh{\rho}_0(X,Y\mid Z)$ and plot the null
distributions in Figure~\ref{boot_distn}(b), from which it can be seen
that the involved estimation of $U$ and $V$ significantly influences the
null distribution of $\wh{\rho}_0(X,Y\mid Z)$. To show the insensitivity
of the choice of the
bandwidth, we set the bandwidths to be $ch_0$, where $c=0.5,$ 1, and 2,
respectively, and $h_0$ is the bandwidth obtained by the rule of thumb. The
estimated kernel density curves of $n\wh{\rho}(X,Y\mid Z)$ and
$n\wh{\rho}_0(X,Y\mid Z)$ based normal distributions with 1000 repetitions,
together with the reference curve, are shown in Figure~\ref{boot_distn}(c) and (d),
from which we can see that the null distributions of  $n\wh{\rho}(X,Y\mid Z)$
almost remain the same for all choices of the bandwidths,
implying that our test is insensitive to
the bandwidth of the nonparametric kernel. However, the null distributions of
$n\wh{\rho}_0(X,Y\mid Z)$ can be dramatically influenced when the bandwidth
changes.

Next, we perform the sensitivity analysis under the alternative hypothesis using Models M2 - M6, which will be listed shortly. We fix the sample size $n=100$ and set the significance level  $\alpha=0.05$. To inspect how the power performance is varied with the choice of the
bandwidth, we set the bandwidths to be $ch_0$, where $h_0$ is the bandwidth obtained by the rule of thumb and increase $c$ from 0.5 to 1.5 step by 0.1. The respective empirical powers are charted in Table \ref{tableh},
from which we can see that the test is the most powerful when $c$ is around 1. Therefore, we advocate  using the rule of thumb (i.e., $c=1$) to decide the bandwidth in practice.

\begin{table}
	\begin{center}
		\caption{\label{tableh} Empirical power of tests of conditional independence for Models M2 - M6 for different bandwidth $ch_0$, where $c$ increase from 0.5 to 1.5. The significance level $\alpha = 0.05$, $n=100$.
		}\setlength\tabcolsep{4pt}
		\vspace{0.25cm}
		\begin{tabular}{cccccccccccc}\hline
			&   0.5 &   0.6 &   0.7 &   0.8 &   0.9 &   1.0 &   1.1 &   1.2 &   1.3 &   1.4 &   1.5 \\\hline
			
			M2 & 1.000 & 1.000 & 1.000 & 1.000 & 1.000 & 1.000 & 1.000 & 1.000 & 1.000 & 1.000 & 1.000 \\
			M3 & 0.957 & 0.962 & 0.98 & 0.977 & 0.975 & 0.971 & 0.972 & 0.968 & 0.955 & 0.960 & 0.956 \\
			M4 & 1.000 & 1.000 & 1.000 & 1.000 & 1.000 & 1.000 & 1.000 & 1.000 & 1.000 & 1.000 & 1.000 \\
			M5 & 0.999 & 1.000 & 0.997 & 0.998 & 0.999 & 0.997 & 0.995 & 0.997 & 0.999 & 0.997 & 0.997 \\
			M6 & 0.999 & 0.999 & 0.999 & 1.000 & 0.999 & 0.999 & 0.996 & 1.000 & 0.999 & 1.000 & 1.000 \\\hline
		\end{tabular}
	\end{center}
\end{table}

We compare our proposed conditional independence test (denoted by ``CIT'')
with some popular nonlinear conditional dependence measure. They are,
respectively,  the conditional distance correlation \citep[denoted by
``CDC'']{wang2015conditional}, conditional mutual information \citep[denoted by
``CMI'']{scutari2010}, and the KCI.test \citep[denoted by
``KCI'']{zhang2011kci}. We conduct 500 replications for each scenario. The
critical values of the CIT are obtained by conducting 1000 simulations. We
first consider the following models with random variable $Z$. In (M1), $X\hDash
Y\mid Z$. This model is designed for examining the empirical Type I error rate.
While (M2)$-$(M6) are designed for examining the power of the proposed test of
conditional independence. Moreover, for M1 - M3, we generate $\wt X_1$, $\wt X_2$ and $Z$ independently from $N(0,1)$. For M4 - M6, we let $Z\sim N(0,1)$, and generate $\wt X_1$, $\wt X_2\sim t_1$ independently to investigate the power of the methods under heavy tailed distributions.

M1:  $X = \wt X_1+Z$, $Y =\wt  X_2+Z$.

M2:   $X=\wt X_1+Z$, $Y=\wt X_1^2+Z$.

M3:   $X=\wt X_1+Z$, $Y=0.5\sin(\pi\wt  X_1)+Z$.

M4:  $X=X_1+Z$, $Y = X_1+X_2 + Z$.

M5:  $X=\sqrt{|X_1Z|}+Z$, $Y = 0.25X_1^2 X_2^2+X_2+Z$.

M6:  $X=\log(|X_1Z|+1)+Z$, $Y = 0.5(X_1^2Z)+X_2+Z$.

The empirical sizes for M1 and powers for the other five models at the
significance levels $\alpha=0.05$ and 0.1 are depicted in Table~\ref{table1}.
In our simulation, we consider two sample sizes $n = 50$ and $100$.
Table~\ref{table1} indicates that the empirical sizes of all the tests are all very close to the level $\alpha$, which means that the Type I error can be controlled very well. As for the empirical power performance of models M2$-$M6, the proposed test outperforms other  tests for both normal data and heavy tailed data, especially when $n=50$.

\begin{table}
	\caption{\label{table1} Empirical size and power of tests of conditional independence when
		$Z$ is random variable at significance levels with $\alpha = 0.05$ and 0.1 and $n=50 $ and 100.}
	\centerline{
		{\begin{tabular}{ccccccccc}
				\hline
				$n$              & $\alpha$          & Test & M1    & M2    & M3    & M4    & M5    & M6    \\ \hline
				\multirow{8}{*}{50}  & \multirow{4}{*}{0.05} & CIT  & 0.056 & 1.000 &   0.572 & 1.000   & 0.954   & 0.888 \\
				&                       & CDC  & 0.050 & 0.886 &  0.338 & 0.837  & 0.881 & 0.473   \\
				&                       & CMI  & 0.048 & 0.380 & 0.070 & 0.829  & 0.898 & 0.448  \\
				&                       & KCI  & 0.038 & 0.884 & 0.250 &  0.191  &  0.048 & 0.010\\[6pt]
				& \multirow{4}{*}{0.1}  & CIT  & 0.098 & 1.000 &  0.712& 1.000  & 0.974  & 0.938\\
				&                       & CDC  & 0.134 & 0.970 &  0.562 &0.934   &  0.930 &  0.642\\
				&                       & CMI  & 0.088 & 0.484 & 0.132&  0.854 &  0.912 & 0.485\\
				&                       & KCI  & 0.088 & 0.968 &  0.344& 0.323  &  0.145 &  0.042\\ \hline
				\multirow{8}{*}{100} & \multirow{4}{*}{0.05} & CIT  & 0.048 & 1.000 &  0.960 & 1.000   &   1.000&  0.997\\
				&                       & CDC  & 0.066 & 0.998 & 0.694 &  0.918  &   0.971 & 0.624 \\
				&                       & CMI  & 0.054 & 0.402 &  0.070 & 0.877  &   0.904&  0.424 \\
				&                       & KCI  & 0.040 & 1.000 &  0.444 & 0.371   &   0.095 & 0.020  \\[6pt]
				& \multirow{4}{*}{0.1}  & CIT  & 0.112 & 1.000 &  0.998 & 1.000  & 1.000  & 0.999\\
				&                       & CDC  & 0.158 & 1.000 &  0.834 &  0.974 &  0.985 & 0.745\\
				&                       & CMI  & 0.098 & 0.496 &  0.124 & 0.902  &  0.926 & 0.455\\
				&                       & KCI  & 0.088 & 1.000 &  0.598 &  0.513 &   0.199& 0.057\\ \hline
			\end{tabular}
	}}
\end{table}

We next examine the finite sample performance of the tests when $Z$ is
two-dimensional random vector, i.e., $\z=(Z_1, Z_2)$. M7 is designed
for examining the size since $X\hDash Y\mid \z$. Five conditional dependent
model M8--M12 are designed to examine the power of the tests. Similar as M1-M6, we generate $\wt X_1$, $\wt X_2$, $Z_1$ and $Z_2$ independently from $N(0,1)$ in each of the following model.

M7:  $X =\wt  X_1+Z_1+Z_2$, $Y =\wt  X_2+Z_1+Z_2$.

M8: $X =\wt  X_1^2+Z_1+Z_2$, $Y = \log(\wt X_1+10)+Z_1+Z_2$.

M9:  $X = \tanh(\wt X_1)+Z_1+Z_2$, $Y = \log(\wt X_1^2+10)+Z_1+Z_2$.

M10:  $X = \wt X_1^2+Z_1+Z_2$, $Y = \log(\wt X_1Z_1+10)+Z_1+Z_2$.

M11:  $X = \wt X_1+Z_1+Z_2$, $Y = \sin(\wt X_1Z_1)+Z_1+Z_2$.

M12:  $X = \log(\wt X_1Z_1+10)+Z_1+Z_2$, $Y = \exp(\wt X_1Z_2)+Z_1+Z_2$.

\begin{table}
	\caption{\label{table2} Empirical size and power of conditional independence tests
		with $\z$ being random vector, level $\alpha$ = 0.05 and 0.1, and sample size
		$n=$ 50 and 100.}
	\centerline{
		{\begin{tabular}{ccccccccc}
				\hline
				$n$              & $\alpha$          & Test & M7    & M8    & M9    & M10    & M11    & M12    \\ \hline
				\multirow{8}{*}{50}  & \multirow{4}{*}{0.05} & CIT  & 0.046 & 0.672 & 0.906 & 0.686 & 0.440 & 0.788 \\
				&                       & CDC  & 0.052 & 0.408 & 0.850 & 0.054 & 0.128 & 0.134 \\
				&                       & CMI  & 0.072 & 0.426 & 0.192 & 0.392 & 0.118 & 0.300 \\
				&                       & KCI  & 0.046 & 0.030 & 0.088 & 0.026 & 0.662 & 0.248 \\[6pt]
				& \multirow{4}{*}{0.1}  & CIT  & 0.092 & 0.792 & 0.948 & 0.798 & 0.582 & 0.874 \\
				&                       & CDC  & 0.126 & 0.670 & 0.976 & 0.194 & 0.298 & 0.264 \\
				&                       & CMI  & 0.120 & 0.506 & 0.292 & 0.500 & 0.210 & 0.386 \\
				&                       & KCI  & 0.098 & 0.092 & 0.190 & 0.074 & 0.820 & 0.492 \\ \hline
				\multirow{8}{*}{100} & \multirow{4}{*}{0.05} & CIT  & 0.048 & 0.936 & 0.998 & 0.936 & 0.664 & 0.988 \\
				&                       & CDC  & 0.084 & 0.890 & 1.000 & 0.920 & 0.972 & 0.306 \\
				&                       & CMI  & 0.044 & 0.412 & 0.164 & 0.392 & 0.126 & 0.300 \\
				&                       & KCI  & 0.044 & 0.038 & 0.172 & 0.028 & 0.990 & 0.358 \\[6pt]
				& \multirow{4}{*}{0.1}  & CIT  & 0.104 & 0.958 & 1.000 & 0.966 & 0.766 & 0.996 \\
				&                       & CDC  & 0.168 & 0.978 & 1.000 & 0.986 & 0.992 & 0.456 \\
				&                       & CMI  & 0.128 & 0.486 & 0.240 & 0.460 & 0.194 & 0.390 \\
				&                       & KCI  & 0.090 & 0.092 & 0.358 & 0.108 & 0.998 & 0.608 \\ \hline
	\end{tabular}}}
\end{table}

{\color{black}
	Lastly, we study the finite sample performance of the tests when $X$, $Y$ and $Z$ are all multivariate. Specifically, $\x=(X_1, X_2)$, $\y=(Y_1,Y_2)$, $\z=(Z_1,Z_2)$. M13 is designed to examine the size of the tests, while M14 -M18 are designed to study the powers. We generate $\wt X_1$, $X_2$, $Y_2$, $Z_1$ and $Z_2$ independently from $N(0,1) $ for each model in M13-M18. 
	
	M13:  $X_1 = \wt X_1+Z_1$, $Y_1= Z_1+Z_2$.
	
	M14: $X_1 = \log(\wt X_1Z_1+100)+Z_1+Z_2$, $Y_1= \exp(\wt X_1Z_1)+Z_1+Z_2$.
	
	M15:   $X = \log(\wt X_1^2+100)+Z_1+Z_2$, $Y_1 = 0.1\wt X_1^3+Z_1+Z_2$.
	
	M16:  $X = \log(\wt X_1*Z_1+100)+Z_1+Z_2$, $Y_1 = 0.5\wt X_1^3 Z_1^3+Z_1+Z_2$.
	
	M17:  $X = 0.1\exp(\wt X_1)+Z_1+Z_2$, $Y_1 = \sin(\wt X_1)+|\wt X_1|+Z_1+Z_2$.
	
	M18:  $X = \tanh(\wt X_1)+Z_1+Z_2$, $Y _1= 0.5\log(\wt X_1^2+100)+0.5X_2+Z_1+Z_2$.

	Simulation results of models M7$-$M12 and models M13$-$M18 are summarized in Tables \ref{table2} and \ref{tableMulti}, respectively, from which it can be seen that the proposed method outperforms all other tests in terms of type I error and power. Furthermore, the numerical results seem to indicates that  when the conditional set is large,
	the conditional mutual information and the kernel based conditional test tend to have relatively low
	power. The conditional distance correlation has high power but suffers huge
	computational burden.
}

\begin{table}
	\caption{\label{tableMulti} Empirical size and power of conditional independence tests
		when $\x$, $\y$ and $\z$ are random vectors. Level $\alpha$ = 0.05 and 0.1, and sample size
		$n=$ 50 and 100.}
	\centerline{
		{\begin{tabular}{ccccccccc}
				\hline
				$n$              & $\alpha$          & Test & M13    & M14    & M15    & M16    & M17   & M18    \\ \hline
				\multirow{8}{*}{50}  & \multirow{4}{*}{0.05} & CIT & 0.05 & 1.000 & 1.000 & 1.000 & 0.363 & 0.986 \\
				&  & CDC & 0.019 & 0.022 & 0.984 & 0.304 & 0.120 & 0.833 \\
				&  & CMI & 0.01 & 0.582 & 0.812 & 0.205 & 0.082 & 0.020 \\
				&  & KCI & 0.036 & 0.036 & 0.047 & 0.042 & 0.052 & 0.688 \\ [6pt]
				& \multirow{4}{*}{0.1}  & CIT & 0.100 & 1.000 & 1.000 & 1.000 & 0.564 & 0.997 \\
				&  & CDC & 0.092 & 0.064 & 1.000 & 0.733 & 0.262 & 0.939 \\
				&  & CMI & 0.028 & 0.6 & 0.915 & 0.294 & 0.170 & 0.054 \\
				&  & KCI & 0.086 & 0.081 & 0.099 & 0.083 & 0.118 & 0.886 \\  \hline
				\multirow{8}{*}{100} & \multirow{4}{*}{0.05} & CIT & 0.026 & 1.000 & 1.000 & 1.000 & 0.873 & 1 \\
				&  & CDC & 0.048 & 0.032 & 1.000 & 0.965 & 0.38 & 0.999 \\
				&  & CMI & 0.004 & 0.498 & 1.000 & 0.211 & 0.218 & 0.036 \\
				&  & KCI & 0.044 & 0.042 & 0.052 & 0.05 & 0.068 & 0.999 \\ [6pt]
				& \multirow{4}{*}{0.1}  & CIT & 0.077 & 1.000 & 1.000 & 1.000 & 0.965 & 1.000 \\
				&  & CDC & 0.127 & 0.13 & 1.000 & 1.000 & 0.567 & 1.000 \\
				&  & CMI & 0.013 & 0.523 & 1.000 & 0.338 & 0.378 & 0.077 \\
				&  & KCI & 0.087 & 0.077 & 0.102 & 0.091 & 0.119 & 1.000 \\   \hline
	\end{tabular}}}
\end{table}

\subsection{Application to causal discovery}\label{app_to_causal}

In this section, we consider a real application of conditional independence test in
causal discovery of directed acyclic graphs. For a directed acyclic graph $G =
(V, E)$, the nodes $V=\{1, 2, \dots, p\}$ corresponds to a random vector $\x =
(X_1,\dots,X_p)\in\mR^p$, and the set of edges $E\subset V\times V$ do not form
any directed cycles. Two vertices $X_1$ and $X_2$ are d-separated by a
subset of vertices $S$ if every path between them is blocked by $S$. One may
refer to \cite{wasserman2013all} for a formal definition. Denote the joint
distribution of $\x$ by $P(\x)$. The joint distribution is said to be
faithful with respect to a graph $G$ if and only if for any $i, j\in
V$, and any subset $S\subset V$, \beqrs X_i \hDash X_j \mid \{X_r: r\in S \}
\Leftrightarrow  \mbox{node } i \mbox{ and node } j \mbox{ are  $d$-separated
	by the set }S. \eeqrs One of the most famous algorithms for recovering the
graphs satisfying the faithfulness assumption is the PC--algorithm
\citep{spirtes2000causation, kalisch2007estimating}. The algorithm could recover
the graph up to its Markov equivalence class, which are sets of graphs
that entail the same set of (conditional) independencies. The performance of
the PC--algorithm relies heavily on the (conditional) independence tests because
small mistakes at the beginning of the algorithm may lead to a totally
different directed acyclic graph \citep{zhang2011kci}. One of the most popular
approach for testing conditional independence is the partial correlation,
under the assumption that the joint distribution $P(\x)$ follows Gaussian
distribution and the nodes relationship is linear
\citep{kalisch2007estimating}. Conditional mutual information
\citep{scutari2010} is another possible option.
\cite{zhang2011kci} proposed a kernel-based conditional independence test for
causal discovery in directed acyclic graphs. In this section, we demonstrate how
the proposed conditional independence index can be applied for causal discovery in real data. Additional simulation results are relegated into the appendix.

\begin{figure}[h!]
	\centerline{
		\begin{tabular}{ccc}
			\psfig{figure=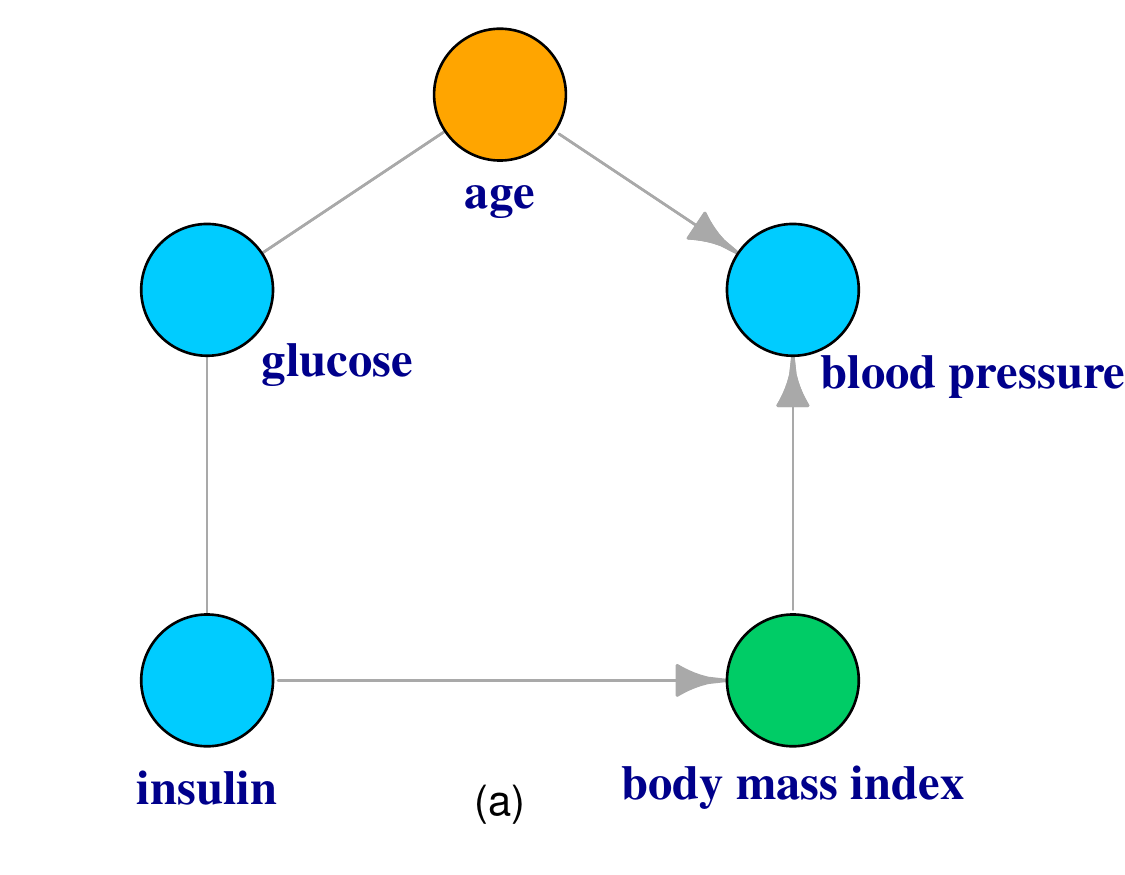,width=3in,height=2.5in,angle=0} &
			\psfig{figure= 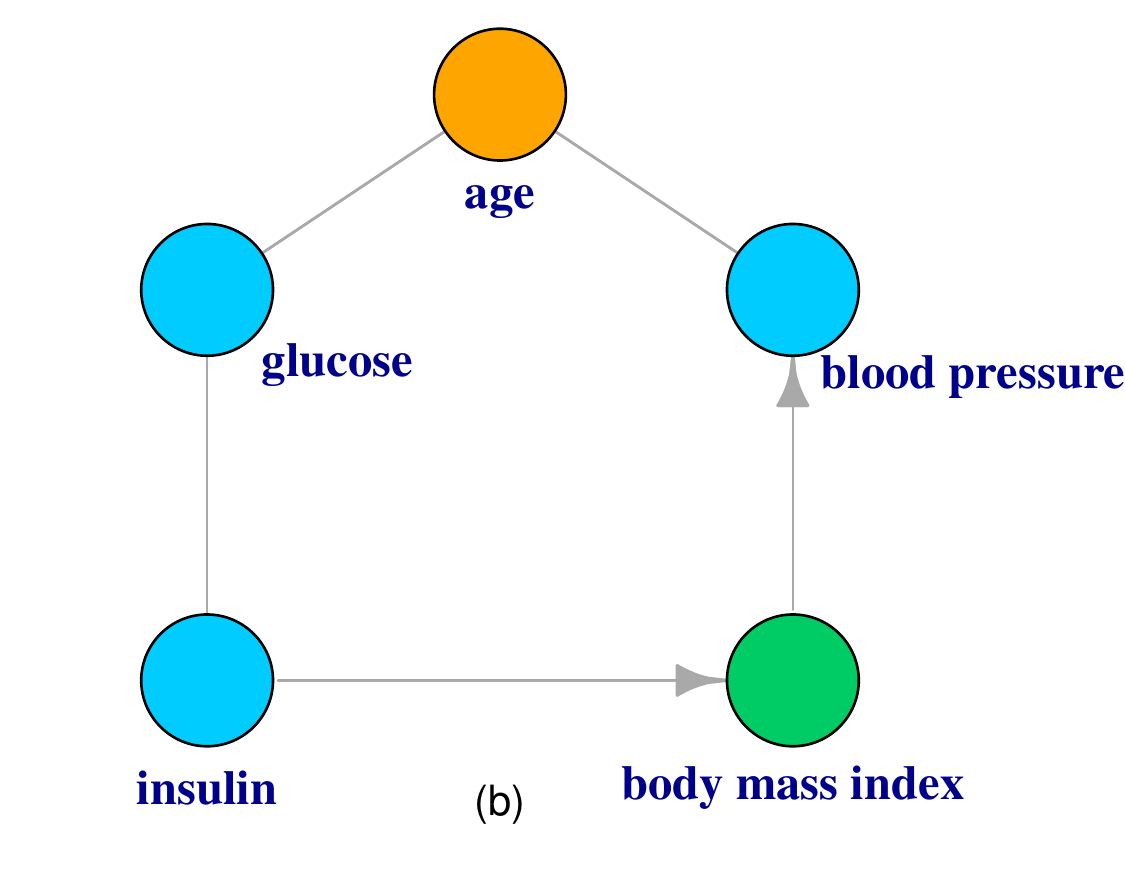,width=3in,height=2.5in,angle=0} \\
			(a) & (b) \\
			\psfig{figure=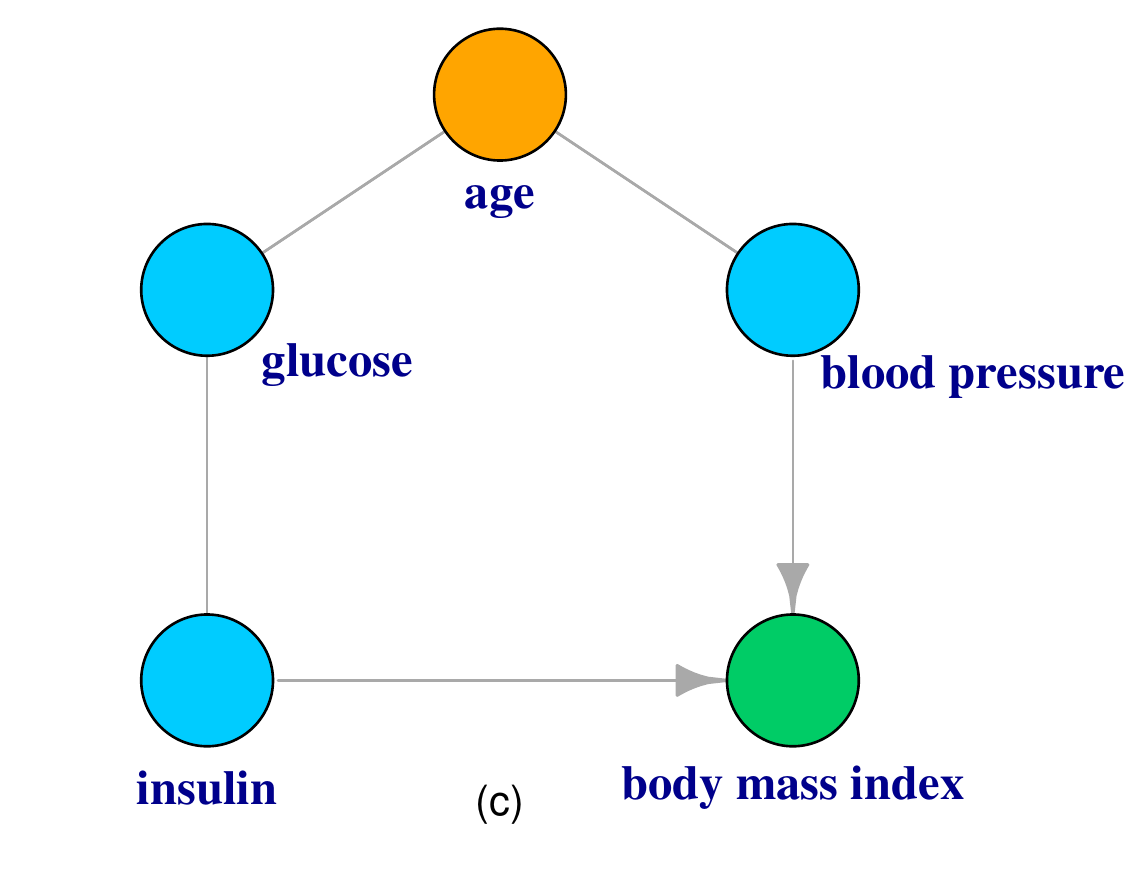,width=3in,height=2.5in,angle=0} &
			\psfig{figure= 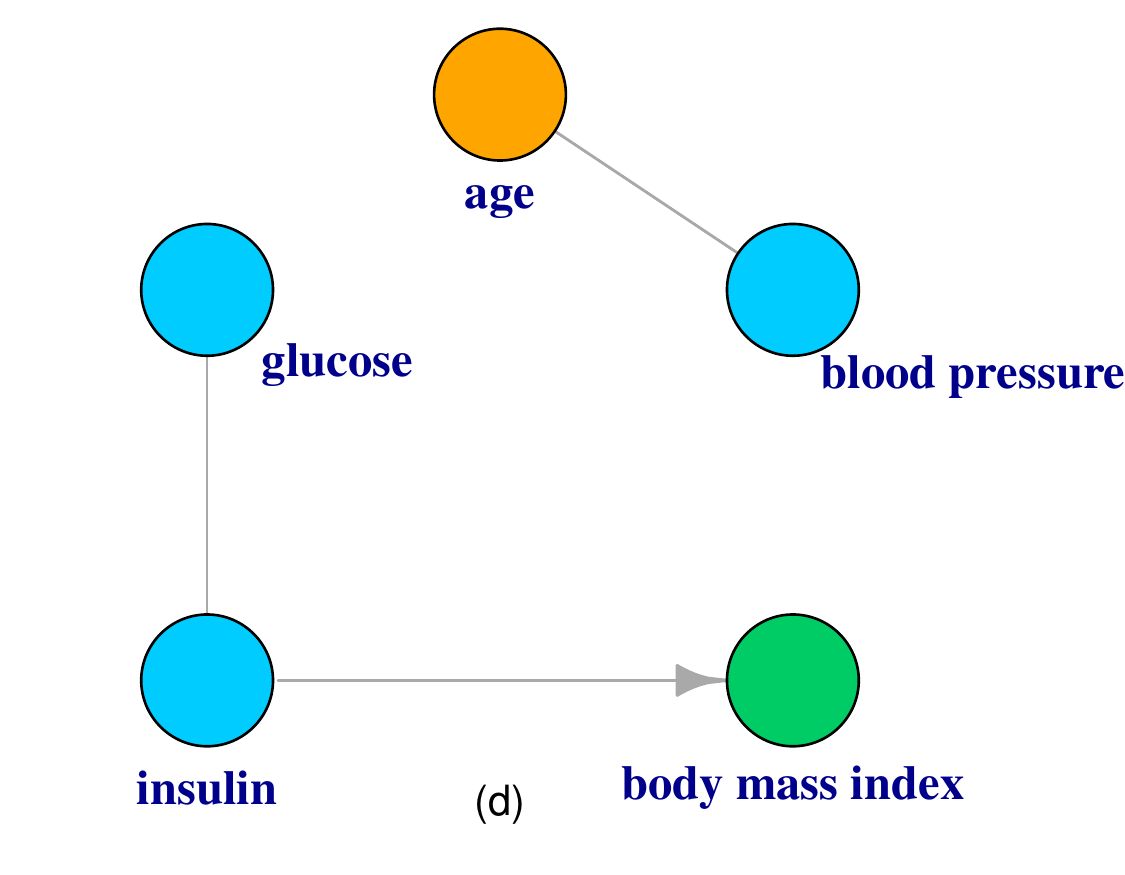,width=3in,height=2.5in,angle=0} \\
			(c) & (d) \\
		\end{tabular}
	}
	\caption{The estimated causal structure of the five
		variables by using the proposed  test in (a), partial correlation in (b),
		conditional mutual information in (c) and kernel based conditional
		independence test in (d). }
	\label{causal}
\end{figure}

\begin{figure}[h!]
	\centerline{
		\begin{tabular}{ccc}
			\psfig{figure=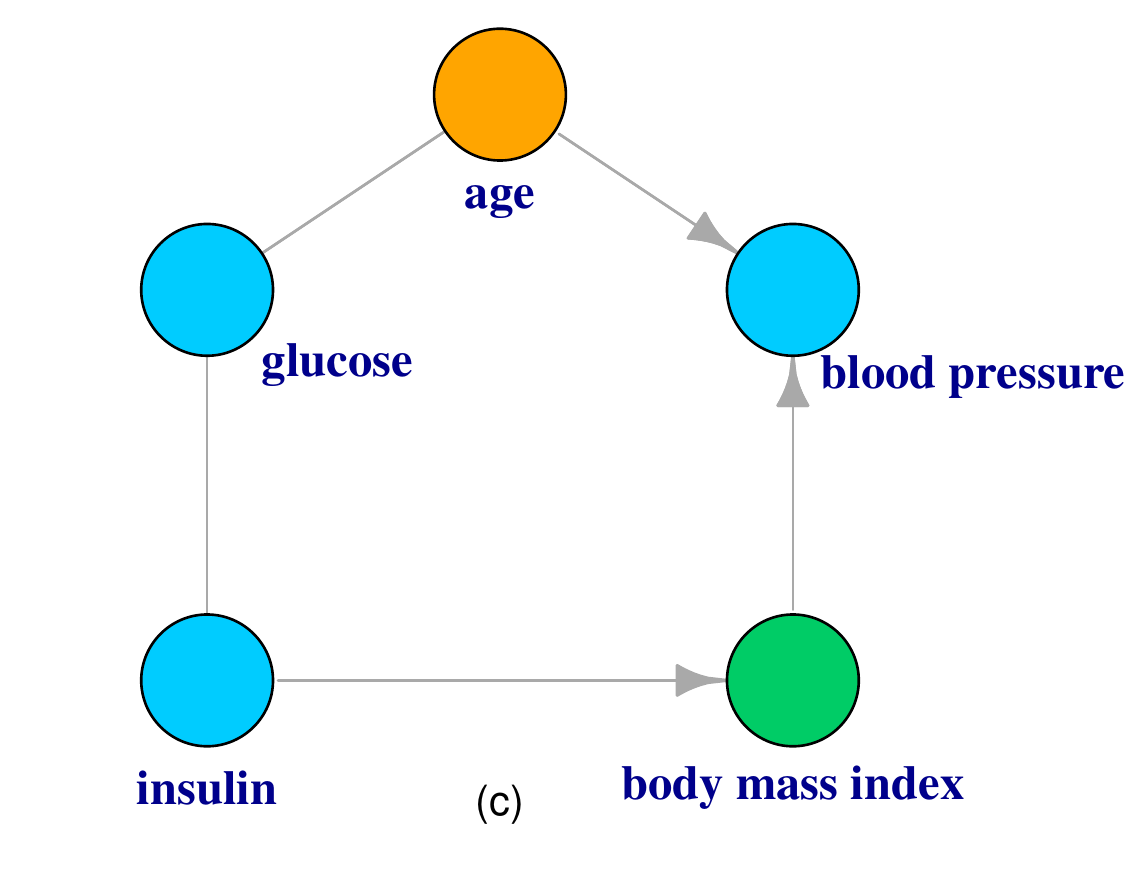,width=3in,height=2.5in,angle=0} &
			\psfig{figure= 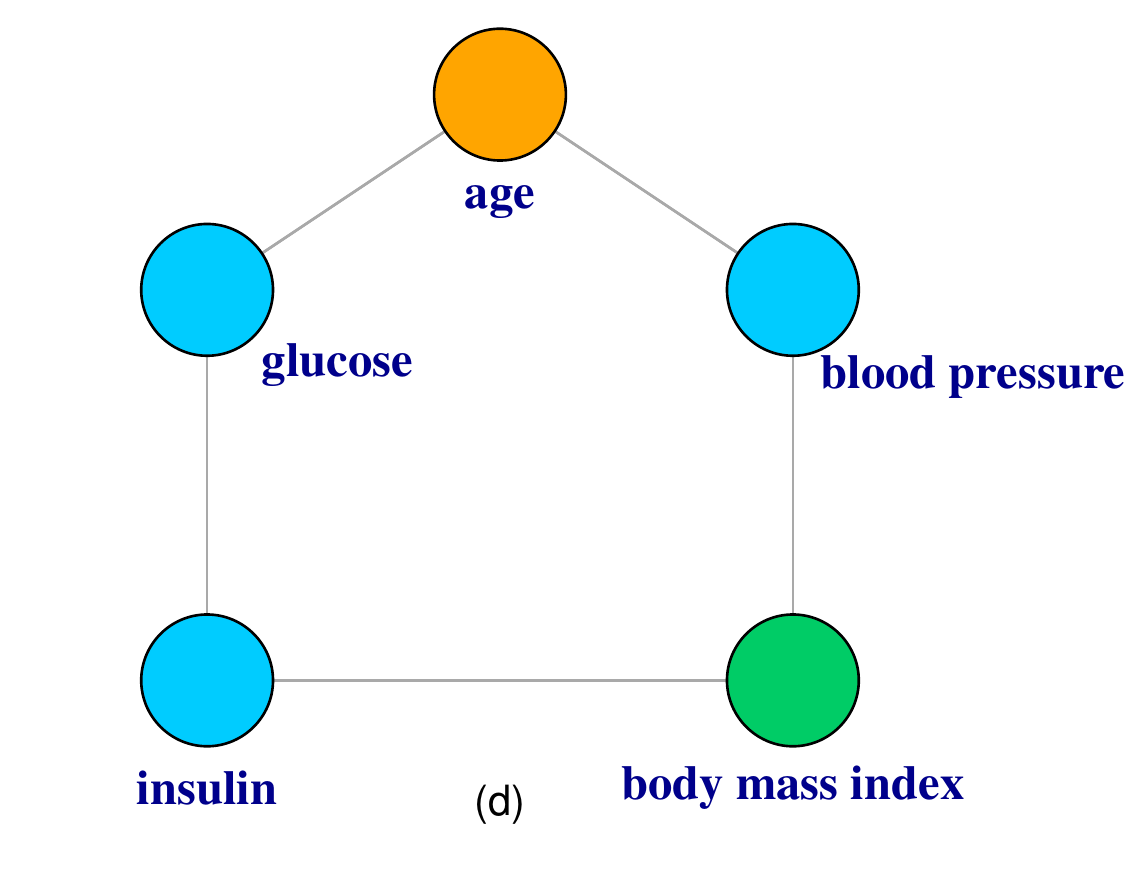,width=3in,height=2.5in,angle=0} \\
			(a) & (b) \\
			\psfig{figure=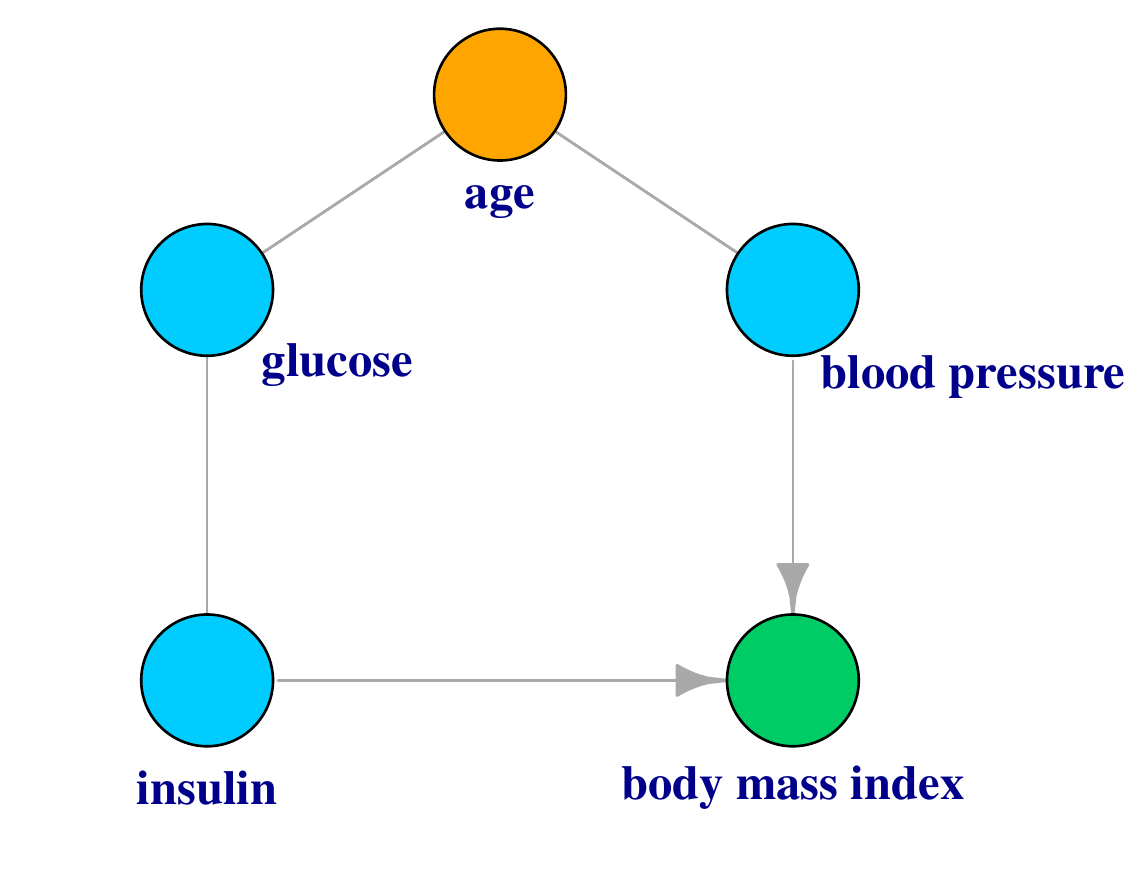,width=3in,height=2.5in,angle=0} &
			\psfig{figure= 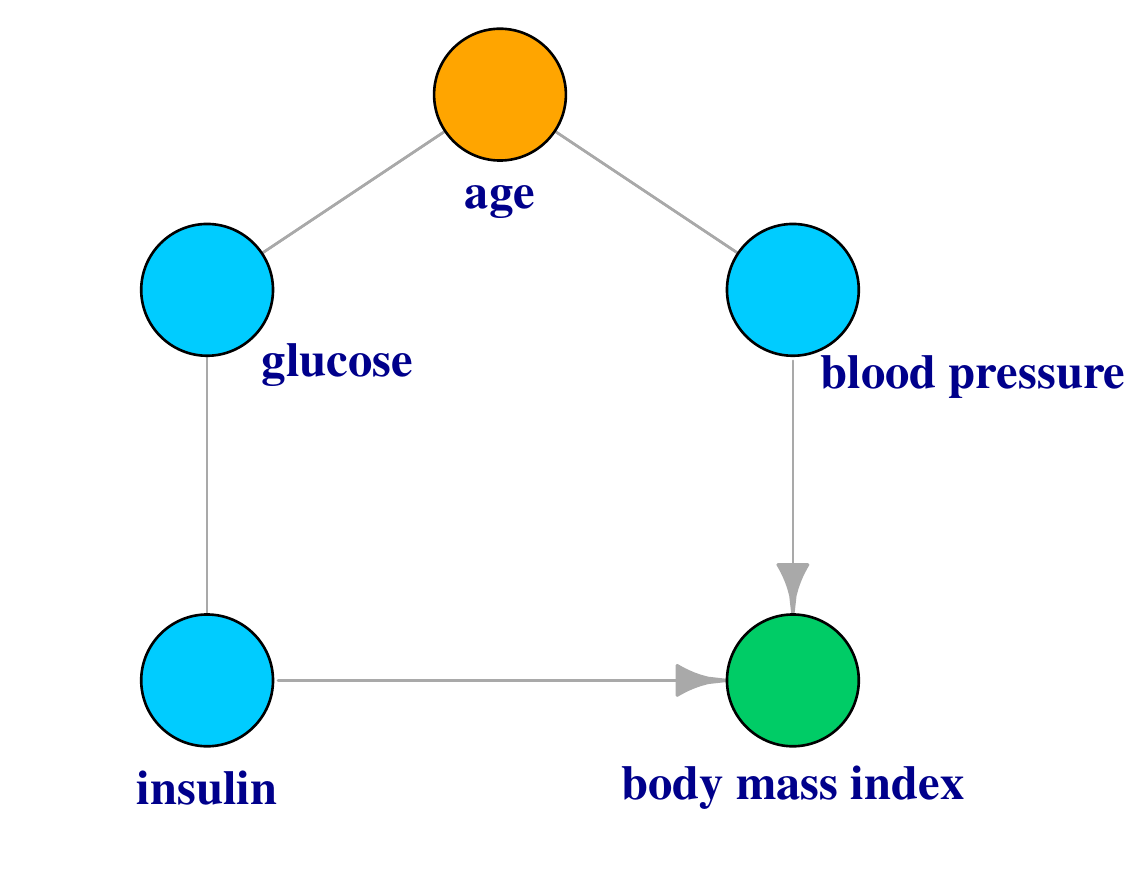,width=3in,height=2.5in,angle=0} \\
			(c) & (d) \\
		\end{tabular}
	}
	\caption{The estimated causal structure of the log-transformed five variables by using the four tests. Refer to the caption of Figure \ref{causal} for the four tests.}
	\label{causal_log}
\end{figure}

We analyze a real data set originally from the National Institute of Diabetes and Digestive and Kidney Diseases \citep{smith1988using}. The dataset consists of serval medical predictor variables for the outcome of diabetes.
We are interested in the causal structural of five variables: age, body mass index, 2-hour serum insulin, plasma glucose concentration and diastolic blood pressure. After removing the missing data, we obtain $n=392$ samples. The PC--algorithm is applied to examine the causal structure of the five variables based on the four different conditional independence measures. We implement the causal algorithms by the R package {\it pcalg} \citep{kalisch2012causal}. The estimated causal structure are shown in Figure \ref{causal}.
The proposed test gives the same estimated graph as the partial correlation,
since the data is approximately normally distributed. To interpret the graph,
note that age is likely to affect the diastolic blood pressure. The plasma glucose concentration level is also likely to be related to age. This is confirmed by the causal findings of (a), (b) and (c) in Figure \ref{causal}. Besides, serum insulin has plausible causal effects on body mass index, and is also related to plasma glucose concentration. The causal relationship between age and blood pressure is not confirmed in part (c),  the test of conditional mutual information. This is not a surprise given the high false positive rate reported in Table \ref{causal1} in the appendix. The kernel based conditional independence is a little conservative and is not able to detect some of the possible edges. To further illustrate
the robustness of the proposed test, we make a logarithm transformation on the data, and apply the same procedure again. The estimated causal structures are reported in Figure \ref{causal_log}.
We observe that the proposed test results in the same estimated structure as the original data, which echos property (4) in Theorem \ref{rho_property}, i.e., the proposed test is invariant with respect to monotone transformations. However, the partial correlation test yields more false positives, since the normality assumption is violated.

\section{Discussions}

In this paper we developed a new index to measure
conditional dependence of random variables and vectors.
The calculation of the estimated index requires
low computational cost. The test of conditional independence
based on the newly proposed index has nontrivial power against all fixed and
local alternatives. The proposed test is distribution free under the null hypothesis,
and is robust to outliers and heavy-tailed data.
Numerical simulations indicate that the proposed test is
more powerful than some existing ones.
The proposed test is further applied to directed acyclic graphs for
causal discovery and shows superior performance.

\section{Technical Proofs}

\subsection{Proof of Proposition \ref{equivalence}}

For $0<u<1$, define quantile function for  $X\mid Z$ as
\[
F_{X\mid Z}^{-1}(u\mid Z=z) = \inf\{x:F_{X\mid Z}(x\mid Z=z)\ge u\}.
\]
Similarly, we can define $F_{Y\mid Z}^{-1}(v\mid Z=z)$, the quantile
function for  $Y\mid Z$, for $0<v<1$.
Since $X$ and $Y$ have continuous conditional distribution functions
for every given value of $Z$,
it follows that when $0<u<1$ and $0<v<1$,
\begin{eqnarray*}
	&&\pr \{F_{X\mid Z}(X\mid Z)\le u, F_{Y\mid Z}(Y\mid Z)\le v\mid Z=z\} \\
	&=& \pr \{X\le F_{X\mid Z}^{-1}(u\mid Z),Y\le  F_{Y\mid Z}^{-1}(v\mid Z)\mid Z=z\}.
\end{eqnarray*}
This implies that $X\hDash Y\mid Z$ is equivalent to $U\hDash V\mid Z$.
In addition, conditional on $Z= z$, $F_{X\mid Z}(X\mid Z= z)$ is
uniformly distributed on $(0,1)$, which does not depend on the
particular value of $ z$, indicating $F_{X\mid Z}(X\mid Z) \hDash Z$.
That is, $U\hDash Z$. Similarly,  $V\hDash Z$.
Thus, the conditional independence $f_{U,V\mid Z}(u,v\mid z) = f_{U\mid Z}(u\mid z)f_{V\mid Z}(v\mid z)$ together with $f_{U\mid Z}(u\mid z)=f_{U}(u)$ and $f_{V\mid Z}(v\mid z) = f_{V}(v)$ implies that
\beqrs
f_{U,V\mid Z}(u,v\mid z) = f_{U}(u)f_{V}(v).
\eeqrs
Thus, $U$, $V$ and $Z$ are mutually independent.

On the other hand, the mutual independence immediately leads to the conditional
independence $U\hDash V\mid Z$. Therefore, the conditional independence
$X\hDash Y\mid Z$ is equivalent to the mutual independence of $U$, $V$ and $Z$.
We next show that the mutual independence of $U$, $V$ and $Z$ is equivalent to mutual
independence of $U$, $V$ and $W$.

Define $
F_{Z}^{-1}(w) = \inf\{z: F(z)\geq w \}
$ for $0<w<1$. If  $U$, $V$ and $Z$ are mutually independent, then
\beqrs
&&\pr(U\leq u, V\leq v, W\leq w)  = \pr\{U\leq u, V\leq v,Z\leq F_{Z}^{-1}(w)\}\\
&=& \pr(U\leq u)\pr( V\leq v)\pr\{Z\leq F_{Z}^{-1}(w)\}= \pr(U\leq u)\pr( V\leq v)\pr(W\leq w)
\eeqrs
holds for all $u,$ $v$ and $w$. On the other hand,
if  $U$, $V$ and $W$ are mutually independent, it follows that
\beqrs
&&\pr(U\leq u, V\leq v, Z\leq z)  = \pr\{U\leq u, V\leq v,W\leq F_{Z}(z)\}\\
&=& \pr(U\leq u)\pr( V\leq v)\pr\{W\leq F_{Z}(z)\}
= \pr(U\leq u)\pr( V\leq v)\pr(Z\leq z),
\eeqrs
holds for all $u,$ $v$ and $z$. Thus, the mutual independence of $U$, $V$ and $Z$ is equivalent to the mutual independence of $U$, $V$ and $W$.
This completes the proof.
\hfill$\fbox{}$

\subsection{Proof of Theorem \ref{rho_property}}

We start with the derivation of
the index $\rho$. $U$, $V$ and $W$ are mutually independent if and
only if \beqr\label{charac_diff} \iiint\l\|\varphi_{U,V,W}(t_1,t_2,t_3) -
\varphi_{U}(t_1)\varphi_{V}(t_2)\varphi_{W}(t_3)\r\|^2\omega(t_1,t_2,t_3)dt_1dt_2dt_3=0,
\eeqr for arbitrary positive weight function $\omega(\cdot)$. We now show that
the proposed index $\rho$  is proportional to the integration in
(\ref{charac_diff}) by choosing $\omega(t_1,t_2,t_3)$ to be the joint
probability density function of three independent and identically distributed
standard Cauchy random variables.

With some calculation and Fubini's theorem, we have
\begin{eqnarray*}
	&&\iiint \l\|\varphi_{U,V,W}(t_1,t_2,t_3) - \varphi_{U}(t_1)\varphi_{V}(t_2)\varphi_{W}(t_3)\r\|^2\omega(t_1,t_2,t_3)dt_1dt_2dt_3\\
	& =& E \iiint e^{it_1(U_1-U_2)+it_2(V_1-V_2)+it_3(W_1-W_2)} \omega(t_1,t_2,t_3)dt_1dt_2dt_3\\
	&& -E \iiint e^{it_1(U_1-U_3)+it_2(V_1-V_4)+it_3(W_1-W_2)} \omega(t_1,t_2,t_3)dt_1dt_2dt_3\\
	&& -E\iiint e^{it_1(U_3-U_1)+it_2(V_4-V_1)+it_3(W_2-W_1)} \omega(t_1,t_2,t_3)dt_1dt_2dt_3\\
	&& +E\iiint e^{it_1(U_1-U_2)+it_2(V_3-V_4)+it_3(W_5-W_6)} \omega(t_1,t_2,t_3)dt_1dt_2dt_3.
\end{eqnarray*}
According to the property of characteristic function for standard Cauchy distribution, we have
\beqrs
\int e^{it(U_1-U_2)} \pi^{-1} (1+t^2)^{-1}dt = e^{-|U_1-U_2|}.
\eeqrs
Then by choosing $\omega(t_1,t_2,t_3) = \pi^{-3} (1+t_1^2)^{-1}(1+t_2^2)^{-1}(1+t_3^2)^{-1}$, i.e., the joint density function of three i.i.d. standard Cauchy distributions, we have
\begin{align}\label{cal_diff}
	\begin{split}
		&\iiint \l\|\varphi_{U,V,W}(t_1,t_2,t_3) - \varphi_{U}(t_1)\varphi_{V}(t_2)\varphi_{W}(t_3)\r\|^2\omega(t_1,t_2,t_3)dt_1dt_2dt_3\\
		& = E e^{-|U_1-U_2|-|V_1-V_2|-|W_1-W_2|}  - 2Ee^{-|U_1-U_3|-|V_1-V_4|-|W_1-W_2|}  \\
		& \hspace{1cm}  + Ee^{- |U_1-U_2|}Ee^{-|V_1-V_2|}Ee^{-|W_1-W_2|}.
	\end{split}
\end{align}
Furthermore, with the fact that $U\hDash W$ and $V\hDash W$, (\ref{cal_diff}) is equal to
\beqrs
E\l\{S_U(U_1,U_2)S_V(V_1,V_2)e^{-|W_1-W_2|}\r\},
\eeqrs
where $S_U(U_1,U_2)$ and $S_V(V_1,V_2)$ are defined as
\beqrs
S_U(U_1,U_2) &=& E\l\{e^{-|U_1-U_2|}+e^{-|U_3-U_4|}- e^{-|U_1-U_3|}-e^{-|U_2-U_3|}\mid (U_1,U_2)\r\},\\
S_V(V_1,V_2) &=& E\l\{e^{-|V_1-V_2|}+e^{-|V_3-V_4|}-
e^{-|V_1-V_3|}-e^{-|V_2-V_3|}\mid (V_1,V_2)\r\}. \eeqrs Now we calculate the
normalization constant $c_0$. It follows by the Cauchy-Schwarz inequality that
\beqrs
&&E\l\{S_U(U_1,U_2)S_V(V_1,V_2)e^{-|W_1-W_2|} \r\} \\
&=& E\l[e^{-|W_1-W_2|} E\l\{S_U(U_1,U_2)S_V(V_1,V_2)\mid (W_1,W_2)\r\}\r]\\
&\le& E\l[e^{-|W_1-W_2|} E^{1/2}\l\{S_U^2(U_1,U_2)\mid (W_1,W_2)\r\} E^{1/2}\l\{ S_V^2(V_1,V_2)\mid (W_1,W_2)\r\}\r]\\
&=& E\l[e^{-|W_1-W_2|} E^{1/2}\l\{S_U^2(U_1,U_2)\r\} E^{1/2}\l\{ S_V^2(V_1,V_2)\r\}\r]\\
&=& 2e^{-1}(6.5e^{-2}-20e^{-1}+6.5)\\
&\defby& c_0^{-1},
\eeqrs
where the equality holds if and only if $
S_U(U_1,U_2)=\lambda\l\{S_V(V_1,V_2)\r\}$
holds with probability 1, where $\lambda\ge0$ (because $E\l\{S_U(U_1,U_2)S_V(V_1,V_2)\r\}$ is nonnegative).
Recall that $U$, $V$ and $W$ are all uniformly distributed on $(0,1)$,
further calculations give us
\beqrs
S_U(U_1,U_2) &=& e^{-|U_1-U_2|} +e^{-U_1}+e^{U_1-1}+e^{-U_2}+e^{U_2-1} + 2e^{-1}-4,\\
S_V(V_1,V_2) &=& e^{-|V_1-V_2|} +e^{-V_1}+e^{V_1-1}+e^{-V_2}+e^{V_2-1} +
2e^{-1}-4. \eeqrs This, together with the normalization constant $c_0$, yield
the expression of the index $\rho(X,Y\mid Z)$. Subsequently, the properties of
the index $\rho(X,Y\mid Z)$ can be established.

(1) $\rho(X,Y\mid Z) \geq 0$ holds obviously. It equals $0$ only when $U$,
$V$, $W$ are mutual independent, which is equivalent to the conditional
independence $X\hDash Y\mid Z$. $\rho(X,Y\mid Z) \leq 1$ holds obviously
according  to the derivation of the index $\rho$.
The equality holds if and only if $S_U(U_1,U_2)=\lambda\l\{S_V(V_1,V_2)\r\}$.
Because $ES_U^2(U_1,U_2)=ES_V^2(V_1,V_2)$, we have $\lambda = 1$. If
$U=V$ or $U+V=1$, it is easy to check $S_U(U_1,U_2) = S_V(V_1,V_2)$. This completes
the proof of Part (1).

(2) This property is trivial according to the definition of $\rho$.

(3) For strictly monotone transformations $m_3(\cdot)$, we have when $m_1(\cdot)$ is strictly increasing,
$U_m=F_{m_1(X)\mid m_3(Z)}\{m_1(X)\mid m_3(Z)\}$ equals $U=F_{X\mid Z}(X\mid Z)$, while when $m_1(\cdot)$ is strictly decreasing, it equals $1-U$.
It can be easily verified that $S_U(U_1,U_2)=S_U(1-U_1,1-U_2)$, then we have $S_{U_m}(U_{m1},U_{m2})=S_U(U_1,U_2)$ no matter whether $m_1(\cdot)$ is strictly increasing or decreasing.
Similarly, let $V_m= F_{m_2(Y)\mid m_3(Z)}\{m_2(Y)\mid m_3(Z)\}$, we obtain that $S_{V_m}(V_{m1},V_{m2})=S_V(V_1,V_2)$.
It is clear that $W_m= F_{m_3(Z)}\{m_3(Z)\}$ equals either $W$ or $1-W$, implying $e^{-|W_{m1}-W_{m2}|}=e^{-|W_1-W_2|}$.
Therefore, we have
\beqrs
E\l\{S_{U_m}(U_{m1},U_{m2})S_{V_m}(V_{m1},V_{m2})e^{-|W_{m1}-W_{m2}|}\r\}=E\l\{S_U(U_1,U_2)S_V(V_1,V_2)e^{-|W_1-W_2|}\r\},
\eeqrs
and it is true that $\rho\l\{m_1(X),m_2(Y)\mid m_3(Z)\r\}= \rho(X,Y\mid Z)$.
\hfill$\fbox{}$

\subsection{Proof of Theorem \ref{rho_null}}

For simplicity, we denote by $g(x)=e^{-|x|}$ and $S_0(x,y) = g\l(x-y\r) +e^{-x}+e^{x-1}+e^{-y}+e^{y-1} + 2e^{-1}-4$. We write $c_0^{-1}\wh{\rho}(X,Y\mid Z)$ as
\beqrs
&& n^{-2}\sum_{i,j} \l\{S_0(\wh U_i,\wh U_j) S_0(\wh V_i,\wh V_j)  g(\wh W_i-\wh W_j)\r\}.
\eeqrs
With Taylor's expansion, when $nh^{4m}\to 0$ and $nh^2/\log^2(n)\to \infty$, we have
\beqrs
&& S_0(\wh U_i,\wh U_j)  \\
&=& \l\{ g'\l( U_i- U_j\r)+e^{ U_i-1} -e^{- U_i} \r\}\Delta U_i  -\l\{ g'\l( U_i- U_j\r)-e^{ U_j-1} +e^{- U_j} \r\}\Delta U_j \\
&+& 2^{-1}\l\{g''\l( U_i- U_j\r)(\Delta U_i-\Delta U_j)^2 +\l(e^{- U_i}+e^{ U_i-1}\r) (\Delta U_i)^2 +\l(e^{- U_j}+e^{ U_j-1}\r) (\Delta U_j)^2\r\} \\
&+& 6^{-1}\l\{g'''\l( U_i- U_j\r)(\Delta U_i-\Delta U_j)^3 +\l(e^{ U_i-1}-e^{- U_i}\r) (\Delta U_i)^3 +\l(e^{ U_j-1}-e^{- U_j}\r) (\Delta U_j)^3\r\} \\
&+&S_0\l( U_i, U_j\r) +o_p(n^{-1})\\
&\defby& S_1(U_i,U_j)+S_2(U_i,U_j)+S_3(U_i,U_j)+S_0(U_i,U_j)+o_p(n^{-1}),
\eeqrs
where $\Delta U_i = \wh U_i - U_i$, and $S_k(U_i,U_j), k=1,2,3$ are defined to be each row in an obvious way. Similarly, we expand $S_0(\wh V_i,\wh V_j)$ as
\beqrs
S_0(\wh V_i,\wh V_j) &=& S_0(V_i,V_j)+S_1(V_i,V_j)+S_2(V_i,V_j)+S_3(V_i,V_j)+o_p(n^{-1}).
\eeqrs
As for $g(\wh W_i-\wh W_j)$, we have
\beqrs
g(\wh W_i-\wh W_j) &=& g(W_i-W_j)+g'(W_i-W_j)(\Delta W_i-\Delta W_j)\\
&&+2^{-1}g''(W_i-W_j)(\Delta W_i-\Delta W_j)^2+o_p(n^{-1}).
\eeqrs
Therefore, it follows that
\beqrs
c_0^{-1}\wh{\rho}(X,Y\mid Z)
&=& 2^{-1} n^{-2}\sum_{i,j}S_0\l( U_i, U_j\r) S_0\l( V_i, V_j\r) g''(W_i-W_j)(\Delta W_i-\Delta W_j)^2\\
&&+ n^{-2}\sum_{i,j}\sum_{0\le k+l\le1} S_k\l( U_i, U_j\r) S_l\l( V_i, V_j\r) g'(W_i-W_j)(\Delta W_i-\Delta W_j)\\
&&+ n^{-2}\sum_{i,j}\sum_{0\le k+l\le3} S_k\l( U_i, U_j\r) S_l\l( V_i, V_j\r)  g\l( W_i- W_j\r)+o_p(n^{-1})\\
&\defby& 2^{-1}Q_1+Q_2+Q_3+o_p(n^{-1}).
\eeqrs
We first show that $Q_1$ is of order $o_p(n^{-1})$. In fact,
\beqrs
Q_1&=&n^{-2}\sum_{i,j} S_0\l( U_i, U_j\r) S_0\l( V_i, V_j\r) g''(W_i-W_j)(\Delta W_i-\Delta W_j)^2\\
&=&n^{-2}\sum_{i,j} S_0\l( U_i, U_j\r) S_0\l( V_i, V_j\r) g''(W_i-W_j)\l\{\l(\Delta W_i\r)^2+\l(\Delta W_j\r)^2\r\}\\
&& +2n^{-2}\sum_{i,j} S_0\l( U_i, U_j\r) S_0\l( V_i, V_j\r) g''(W_i-W_j)(W_i\Delta W_j+W_j\Delta W_i)\\
&& -2n^{-2}\sum_{i,j} S_0\l( U_i, U_j\r) S_0\l( V_i, V_j\r) g''(W_i-W_j)(\wh W_i\wh W_j-W_i W_j)\\
&\defby& Q_{1,1}+Q_{1,2}+Q_{1,3}.
\eeqrs
Under the null hypothesis, $U$, $V$ and $W$ are mutually independent,  it is easy to verify that $E\l\{S_0\l( U_i, U_j\r)\mid (U_i,V_i,W_i,V_j,W_j)\r\}=0$, and hence
\beqrs
E\l\{S_0\l( U_i, U_j\r) S_0\l( V_i, V_j\r) g''(W_i-W_j)\mid (U_i,V_i,W_i)\r\}=0.
\eeqrs
Then for each fixed $i$, we have
\beqrs
n^{-1}\sum_{j} S_0\l( U_i, U_j\r) S_0\l( V_i, V_j\r) g''(W_i-W_j)=O_p(n^{-1/2}).
\eeqrs
Thus, $Q_{1,1}$ is clearly of order $o_p(n^{-1})$ because $\l(\Delta W_i\r)^2=o_p(n^{-1/2})$.
Now we deal with $Q_{1,2}$.
\beqrs
&n^{-2}\sum_{i,j} S_0\l( U_i, U_j\r) S_0\l( V_i, V_j\r) g''(W_i-W_j)W_i\Delta W_j\\
&= n^{-3}\sum_{i,j,k} S_0\l( U_i, U_j\r) S_0\l( V_i, V_j\r) g''(W_i-W_j)W_i\l\{\indic (W_k\le W_j)-W_j\r\}. \label{ustat1}
\eeqrs
Under the null hypothesis,  because $E\l\{S_0\l( U_i, U_j\r)\mid (U_i,V_i,W_i,V_j,W_j)\r\}=0$, $W$ is uniformly distributed, we have $E\{\indic (W_k\le W_j)\mid W_j\}=W_j$. Thus, the corresponding U-statistic of the equation above is second order degenerate. In addition,
when any two of $i,j,k$ are identical, we have
\beqrs
E \l[S_0\l( U_i, U_j\r) S_0\l( V_i, V_j\r) g''(W_i-W_j)W_i\l\{\indic (W_k\le W_j)-W_j\r\}\r] =0.
\eeqrs
Then  the summations associated with any two of the $i,j,k$ are identical
is of order $o_p(1)$. Therefore, $Q_{1,2}=o_p(n^{-1})$.
It remains to deal with $Q_{1,3}$.
Similarly, the corresponding U-statistic of
\beqrs
n^{-4}\sum_{i,j,k,l} S_0\l( U_i, U_j\r) S_0\l( V_i, V_j\r) g''(W_i-W_j)\l\{\indic(W_k\le W_i)\indic(W_l\le W_j)-W_i W_j\r\}
\eeqrs
is second order degenerate and hence we obtain that $Q_{1,3}=o_p(n^{-1})$.

Next, we show $Q_2=o_p(n^{-1})$.
Recall that
\beqrs
Q_2&=& n^{-2}\sum_{i,j} S_0\l( U_i, U_j\r) S_0\l( V_i, V_j\r) g'(W_i-W_j)(\Delta W_i-\Delta W_j)\\
&&+2n^{-2}\sum_{i,j} S_1\l( U_i, U_j\r) S_0\l( V_i, V_j\r) g'(W_i-W_j)\Delta W_i\\
&&+2n^{-2}\sum_{i,j} S_0\l( U_i, U_j\r) S_1\l( V_i, V_j\r) g'(W_i-W_j)\Delta W_i\\
&\defby&Q_{2,1}+2Q_{2,2}+2Q_{2,3}.
\eeqrs
Similar to dealing with $Q_{1,2}$, we have $Q_{2,1}=o_p(n^{-1})$. We now evaluate $Q_{2,2}$.
\beqrs
Q_{2,2}&=& n^{-2}\sum_{i,j}\l\{ g'\l( U_i- U_j\r)+e^{ U_i-1} -e^{- U_i} \r\}\Delta U_i S_{0}\l( V_i, V_j\r) g'(W_i-W_j)\Delta W_i\\
&&-n^{-2}\sum_{i,j}\l\{ g'\l( U_i- U_j\r)-e^{ U_j-1} +e^{- U_j} \r\}\Delta U_j  S_{0}\l( V_i, V_j\r) g'(W_i-W_j)\Delta W_i.
\eeqrs
Because under the null hypothesis, $E\l\{S_{0}\l( V_i, V_j\r)\mid (U_i,V_i,W_i,U_j,W_j)\r\}=0$, it follows that for each $i$,
\beqrs
n^{-1}\sum_{j}\l\{ g'\l( U_i- U_j\r)+e^{ U_i-1} -e^{- U_i} \r\} S_{0}\l( V_i, V_j\r) g'(W_i-W_j) =O_p(n^{-1/2}),
\eeqrs
and the first term of $Q_{2,2}$ is of order $o_p(n^{-1})$ because $\Delta W_i=O_p(n^{-1/2})$ and $\Delta U_i=o_p(1)$. In addition, for each $j$,
\beqrs
n^{-2}\sum_{i,k}\l\{ g'\l( U_i- U_j\r)-e^{ U_j-1} +e^{- U_j} \r\} S_{0}\l( V_i, V_j\r) g'(W_i-W_j)\l\{\indic(W_k\le W_i)-W_i\r\}
\eeqrs
is degenerate and hence the second term of $Q_{2,2}$ is also of order $o_p(n^{-1})$ because $\Delta U_j=o_p(1)$, indicating $Q_{2,2}=o_p(n^{-1})$.
Similarly, we have $Q_{2,3}=o_p(n^{-1})$. Thus it follows that $Q_2=o_p(n^{-1})$.

Finally, we show that $Q_3=n^{-2}\sum_{i,j} S_0\l( U_i, U_j\r) S_0\l( V_i, V_j\r)  g\l( W_i- W_j\r)+o_p(n^{-1})$. Or equivalently, we show that $Q_{3,1}, Q_{3,2}$ and $Q_{3,3}$ are all of order $o_p(n^{-1})$, where
\beqrs
Q_{3,1} &\defby&  n^{-2}\sum_{i,j}\sum_{k+l=1} S_k\l( U_i, U_j\r) S_l\l( V_i, V_j\r)  g\l( W_i- W_j\r),\\
Q_{3,2}&\defby& n^{-2}\sum_{i,j}\sum_{k+l=2} S_k\l( U_i, U_j\r) S_l\l( V_i, V_j\r)  g\l( W_i- W_j\r),\\
Q_{3,3}&\defby& n^{-2}\sum_{i,j}\sum_{k+l=3} S_k\l( U_i, U_j\r) S_l\l( V_i, V_j\r)  g\l( W_i- W_j\r).
\eeqrs
We first show that $Q_{3,1}\defby Q_{3,1,1}+Q_{3,1,2}=o_p(n^{-1})$, where
\beqrs
Q_{3,1,1} &=& n^{-2}\sum_{i,j} S_1\l( U_i, U_j\r) S_0\l( V_i, V_j\r)  g\l( W_i- W_j\r),\\
Q_{3,1,2}&=&n^{-2}\sum_{i,j} S_0\l( U_i, U_j\r) S_1\l( V_i, V_j\r)  g\l( W_i- W_j\r).
\eeqrs
Without loss of generality, we only show that $Q_{3,1,1} = o_p(n^{-1})$. Calculate
\beqrs
S_1\l( U_i, U_j\r) &=&  \l\{ g'\l( U_i- U_j\r)+e^{ U_i-1} -e^{- U_i} \r\}\Delta U_i  -\l\{ g'\l( U_i- U_j\r)-e^{ U_j-1} +e^{- U_j} \r\}\Delta U_j,\\
\Delta U_i &=& n^{-1}\sum_{k=1}^n \l[\frac{K_h(Z_k-Z_i)\indic(X_k\le X_i)}{ f(Z_i)}- U_i-\frac{U_i\l\{K_h(Z_k-Z_i)-f(Z_i)\r\}}{f(Z_i)}\r]\\
&&\hspace{2cm}+O_p(h^{2m}+n^{-1}h^{-1}\log^2 n).
\eeqrs
Thus, when $nh^{4m}\to 0$ and $nh^2/\log^2(n)\to \infty$,
\beqrs
Q_{3,1,1}&=&2n^{-2}\sum_{i,j} \l\{ g'\l( U_i- U_j\r)+e^{ U_i-1} -e^{- U_i} \r\} S_0\l( V_i, V_j\r)  g\l( W_i- W_j\r)\Delta U_i \\
&=&2n^{-3}\sum_{i,j,k} \bigg( \l\{ g'\l( U_i- U_j\r)+e^{ U_i-1} -e^{- U_i} \r\} S_0\l( V_i, V_j\r)  g\l( W_i- W_j\r)\\
&&\cdot\l[\frac{K_h(Z_k-Z_i)\indic(X_k\le X_i)}{ f(Z_i)}- U_i-\frac{U_i\l\{K_h(Z_k-Z_i)-f(Z_i)\r\}}{f(Z_i)}\r]\bigg)+o_p(n^{-1})\\
&=&\frac{2}{n(n-1)}\sum_{i\neq j} \bigg( \l\{ g'\l( U_i- U_j\r)+e^{ U_i-1} -e^{- U_i} \r\} S_0\l( V_i, V_j\r)  g\l( W_i- W_j\r)\\
&&\cdot\l[\frac{E\l\{K_h(Z_k-Z_i)\indic(X_k\le X_i)-U_iK_h(Z_k-Z_i)\mid(X_i,Z_i)\r\}}{ f(Z_i)}\r]\bigg)+o_p(n^{-1}),
\eeqrs
where the last equality holds due to equations (2)-(3) of section 5.3.4 in \cite{serfling2009approximation} and the fact that $\var\{K_h(Z_i-Z_j)\}=O(h^{-1})$.
Therefore, since
\beqrs
\sup_{X_i,Z_i}\big|E\l\{K_h(Z_k-Z_i)\indic(X_k\le X_i)-U_iK_h(Z_k-Z_i)\mid(X_i,Z_i)\r\}\big|=O(h^m),
\eeqrs
when the $(m-1)$th derivatives of $F_{X\mid Z}(x\mid z)f_Z(z)$ and $f_Z(z)$ with respect to $z$ are locally Lipschitz-continuous, $Q_{3,1,1}$ is clearly of order $o_p(n^{-1})$ by noting that the summation in the last display is degenerate.

Next, we consider $Q_{3,2}$, where
\beqrs
Q_{3,2} &=& n^{-2}\sum_{i,j} S_1\l( U_i, U_j\r) S_1\l( V_i, V_j\r)  g\l( W_i- W_j\r) \\
&+&n^{-2}\sum_{i,j} S_2\l( U_i, U_j\r) S_0\l( V_i, V_j\r)  g\l( W_i- W_j\r)\\
&+& n^{-2}\sum_{i,j} S_0\l( U_i, U_j\r) S_2\l( V_i, V_j\r)  g\l( W_i- W_j\r) \\
&\defby& Q_{3,2,1} +Q_{3,2,2}+Q_{3,2,3}.
\eeqrs
We first show that $Q_{3,2,1}=o_p(n^{-1})$. It follows that
\beqrs
Q_{3,2,1} &=& 2n^{-2}\sum_{i,j}\bigg[\l\{ g'\l( U_i- U_j\r)+e^{ U_i-1} -e^{- U_i} \r\}\l\{ g'\l( V_i- V_j\r)+e^{ V_i-1} -e^{-V_i} \r\}\\
&&\cdot g\l( W_i- W_j\r) \Delta U_i\Delta V_i\bigg] +2n^{-2}\sum_{i,j}\bigg[\l\{ g'\l( U_i- U_j\r)+e^{ U_i-1} -e^{- U_i} \r\}\\
&&\cdot \l\{ g'\l( V_i- V_j\r)-e^{ V_j-1} +e^{- V_j} \r\}g\l( W_i- W_j\r) \Delta U_i\Delta V_j \bigg]\\
&\defby& Q_{3,2,1,1}+Q_{3,2,1,2}.
\eeqrs
Because $E\{g'\l( U_i- U_j\r)+e^{ U_i-1} -e^{- U_i}\mid (U_i,V_i,W_i,V_j,W_j)\} = 0$. For each $i$,
\beqrs
n^{-1}\sum_{j=1}^n\bigg[\l\{ g'\l( U_i- U_j\r)+e^{ U_i-1} -e^{- U_i} \r\}\l\{ g'\l( V_i- V_j\r)+e^{ V_i-1} -e^{-V_i} \r\}g\l( W_i- W_j\r) \bigg]
\eeqrs
is of order $O_p(n^{-1/2})$. Then $Q_{3,2,1,1}=o_p(n^{-1})$ because $\Delta U_i\Delta V_i=o_p(n^{-1/2})$.
For $Q_{3,2,1,2}$, $-\Delta U_i\Delta V_j  = (U_i\Delta V_j+U_j\Delta V_i)+(\wh U_i\wh V_j -U_i V_j)$. By expanding the $\wh U_i$, $\wh U_j$, $\wh V_i$, $\wh V_j $ in $Q_{3,2,1,2}$ as U statistics and apply the same technique as showing $Q_{3,1,1} = o_p(n^{-1})$ and $Q_{1,3}=o_p(n^{-1})$, it follows immediately that $Q_{3,2,1,2}$ is of order $o_p(n^{-1})$. Thus $Q_{3,2,1}$ is of order $o_p(n^{-1})$.

For $Q_{3,2,2}$ and $Q_{3,2,3}$, we only show that $Q_{3,2,2}$ is of order $o_p({n^{-1}})$, for simplicity.
Recall that $S_2\l( U_i, U_j\r)$ is defined as
\beqrs
S_2\l( U_i, U_j\r)&=& 2^{-1}\bigg[g''\l( U_i- U_j\r)(\Delta U_i)^2+g''\l( U_i- U_j\r)(\Delta U_j)^2 +\l(e^{- U_i}+e^{ U_i-1}\r) (\Delta U_i)^2\\
&& \hspace{2cm} +\l(e^{- U_j}+e^{ U_j-1}\r) (\Delta U_j)^2-2g''\l( U_i- U_j\r)\Delta U_i\Delta U_j\bigg].
\eeqrs
The summations associated with either $(\Delta U_i)^2$ or $(\Delta U_j)^2$ are of order $o_p(n^{-1})$ following similar reasons as showing $Q_{3,2,1,1} = o_p(n^{-1})$, and that associated with $\Delta U_i\Delta U_j$ are of order $o_p(n^{-1})$ similar to dealing with $Q_{3,2,1,2}$.
As a result, $Q_{3,2}$ is of order $o_p(n^{-1})$.

For $Q_{3,3}$, we have
\beqrs
Q_{3,3} = n^{-2}\sum_{i,j}\sum_{k=1}^4 S_{4-k}\l( U_i, U_j\r) S_{k-1}\l( V_i, V_j\r)  g\l( W_i- W_j\r)\defby \sum_{k=1}^4Q_{3,3,k}.
\eeqrs
We only show that $Q_{3,3,1}=o_p(n^{-1})$ because the other terms are similar. Calculate
\beqrs
&&6S_3\l( U_i, U_j\r)\\
&=&g'''\l( U_i- U_j\r)\l\{(\Delta U_i)^3-3(\Delta U_i)^2\Delta U_j+3\Delta U_i(\Delta U_j)^2-(\Delta U_j)^3\r\} \\
&&\hspace{3cm}+\l(e^{ U_i-1}-e^{- U_i}\r) (\Delta U_i)^3 +\l(e^{ U_j-1}-e^{- U_j}\r) (\Delta U_j)^3.
\eeqrs
Then the summations associated with either $(\Delta U_i)^3$ or $(\Delta U_j)^3$ are of order $o_p(n^{-1})$ similar to dealing with $Q_{3,2,1,1}$, and that associated with $\Delta U_i(\Delta U_j)^2$ or $(\Delta U_i)^2\Delta U_j$ are of order $o_p(n^{-1})$ similar to the second term of $Q_{2,2}$.

To sum up, we have shown that
\beqrs
c_0^{-1}\wh{\rho}(X,Y\mid Z)=n^{-2}\sum_{i,j} S_0\l( U_i, U_j\r) S_0\l( V_i, V_j\r)  g\l( W_i- W_j\r)+o_p(n^{-1}),
\eeqrs
where the right hand side is essentially a first order degenerate V-statistics. Thus by applying Theorem 6.4.1.B of \cite{serfling2009approximation},
\beqrs
n\wh{\rho}(X,Y\mid Z)\xrightarrow{d} c_0\sum_{j=1}^{\infty}\lambda_j\chi_{j}^{2}(1)
\eeqrs
where $\chi_{j}^2(1)$, $j=1, 2, \dots$ are independent $\chi^2(1)$ random variables, and $\lambda_j$, $j=1, 2, \dots$ are the  corresponding eigenvalues of $h(u,v,w;u',v',w')$. It is worth mentioning that the kernel is positive definite and hence all the $\lambda_j$s are positive. Therefore, the proof is completed.
\hfill$\fbox{}$

\subsection{Proof of Theorem \ref{rho_bootstrap}}
Since we generate $\{U_i^*, V_i^*, W_i^*\}$, $i=1,\dots, n$ independently from uniform distribution, it is quite straightforward that $U^*, V^*$ and $W^*$ are mutually independent.
In addition, we can write $\wh{\rho}^*$ as
\beqrs
\wh{\rho}^*=n^{-2}\sum_{i,j}c_0 S_0\l( U_i^*, U_j^*\r) S_0\l( V_i^*, V_j^*\r)  g\l( W_i^*- W_j^*\r),
\eeqrs
which clearly converges in distribution to
$c_0\sum_{j=1}^{\infty}\wt\lambda_j\chi_{j}^{2}(1)$,
where $\chi_{j}^2(1)$, $j=1, 2, \dots$ are independent $\chi^2(1)$ random variables, and $\wt\lambda_j$, $j=1, 2, \dots$ are the eigenvalues of $h(u,v,w;u',v',w')$, implying $\wt\lambda_j=\lambda_j$, for $j=1, 2, \dots$, and hence the proof is completed.

\hfill$\fbox{}$

\subsection{Proof of Theorem \ref{rho_power}}
We use the same notation as the proof in Theorem \ref{rho_null}. With Taylor's
expansion, when $nh^{4m}\to 0$ and $nh^2/\log^2(n)\to \infty$, we have
\beqrs
S_0(\wh U_i,\wh U_j)&=&S_0(U_i,U_j)+S_1(U_i,U_j)+o_p(n^{-1/2}),\\
S_0(\wh V_i,\wh V_j)&=&S_0(V_i,V_j)+S_1(V_i,V_j)+o_p(n^{-1/2}),\\
g(\wh W_i-\wh W_j) &=& g(W_i-W_j)+g'(W_i-W_j)(\Delta W_i-\Delta W_j)+o_p(n^{-1/2}).
\eeqrs
Therefore, we have
\beqrs
c_0^{-1}\wh{\rho}(X,Y\mid Z)
&=&n^{-2}\sum_{i,j} S_0\l( U_i, U_j\r) S_0\l( V_i, V_j\r)  g\l( W_i- W_j\r)\\
&&+n^{-2}\sum_{i,j}S_0\l( U_i, U_j\r) S_0\l( V_i, V_j\r) g'(W_i-W_j)(\Delta W_i-\Delta W_j)\\
&& +n^{-2}\sum_{i,j}S_1\l( U_i, U_j\r) S_0\l( V_i, V_j\r) g(W_i-W_j)\\
&& +n^{-2}\sum_{i,j}S_0\l( U_i, U_j\r) S_1\l( V_i, V_j\r) g(W_i-W_j)+o_p(n^{-1/2})\\
&\defby& P_1+P_2+P_3+P_4+o_p(n^{-1/2}).
\eeqrs
We deal with the four terms, respectively. For $P_1$, by applying
Lemma 5.7.3 and equation (2) in section 5.3.1 of \cite{serfling2009approximation}, we have
\beqr\nonumber
P_1 -c_0^{-1}\rho(X,Y\mid Z)
&=& 2n^{-1} \sum_{i=1}^{n}E\left[ \l\{ S_0\l( U_i, U\r) S_0\l( V_i, V\r)  g\l( W_i- W\r)  \r\} \mid(U_i,V_i,W_i)\right] \\\nonumber
&& \hspace{3cm}-2c_0^{-1}\rho(X,Y\mid Z)+o_p(n^{-1/2}) \\\label{P1i}
&\defby& 2n^{-1}\sum_{i=1}^{n} \l\{P_{1, i} -c_0^{-1}\rho(X,Y\mid Z)\r\}+o_p(n^{-1/2}).
\eeqr

Next, we deal with $P_2$.
Recall that
\beqrs
P_2&=&n^{-2}\sum_{i,j}S_0\l( U_i, U_j\r) S_0\l( V_i, V_j\r) g'(W_i-W_j)(\Delta W_i-\Delta W_j)\\
&=&2n^{-3}\sum_{i,j,k}S_0\l( U_i, U_j\r) S_0\l( V_i, V_j\r) g'(W_i-W_j)\l\{I(W_k\leq W_i) - W_i\r\}.
\eeqrs
By applying Lemma 5.7.3 and equation (2) in section 5.3.1 of \cite{serfling2009approximation} again, we can obtain that
\beqr\nonumber
P_2&=&2n^{-1}\sum_{i=1}^n E\l[ S_0( U,  U') S_0( V,  V')g'(W- W')\l\{I(W_i\leq W) - W\r\}\mid W_i\r]+o_p(n^{-1/2})\\\label{P2i}
&\defby& 2n^{-1}\sum_{i=1}^{n} P_{2, i}+o_p(n^{-1/2}),
\eeqr
where $( U',V',W')$ is an independent copy of $(U,V,W)$.

It remains to deal with $P_3$ and $P_4$.
$P_{3}$ equals
\beqrs
P_{3}&=&2n^{-3}\sum_{i,j,k}\bigg[\l\{ g'\l( U_i- U_j\r)+e^{ U_i-1} -e^{- U_i} \r\} S_0\l( V_i, V_j\r) g(W_i-W_j)\\
&&\hspace{2cm}\cdot\l\{\frac{K_h(Z_k-Z_i)\indic(X_k\le X_i)-U_iK_h(Z_k-Z_i)}{f(Z_i)}\r\}\bigg]+o_p(n^{-1/2}).
\eeqrs
By definition, we have $V\hDash W$ and hence it can be verified that
\beqrs
E\l\{S_0\l( V_i, V_j\r) g(W_i-W_j)\mid V_i,W_i\r\}=0.
\eeqrs
Denote $P_{3}^{k,i} =\{K_h(Z_k-Z_i)\indic(X_k\le X_i)-U_iK_h(Z_k-Z_i)\}/f_Z(Z_i)$. Thus,
\beqrs
&&E\l\{ \l(e^{ U_i-1} -e^{- U_i}\r)S_0\l( V_i, V_j\r) g(W_i-W_j)P_{3}^{k,i}\mid X_k, Z_k \r\}  \\
&=&E\l\{ \l(e^{ U_i-1} -e^{- U_i}\r)S_0\l( V_i, V_j\r) g(W_i-W_j)P_{3}^{k,i}\mid X_k, Z_k , U_i\r\}  = 0.
\eeqrs
Thus when $nh^{2m}\to0$ and $nh\to\infty$,
we have
\beqr\nonumber
P_{3}&=&2n^{-1}\sum_{k=1}^n E\l[\l\{I(X\ge X_k)-U\r\} g'(U- U')S_0(V,V')\r.\\\nonumber
&&\l.\hspace{2cm}\cdot g(W-W')\mid X_k, Z_k\r]+o_p(n^{-1/2})\\\label{P3i}
&\defby& 2n^{-1}\sum_{i=1}^n P_{3,i}+o_p(n^{-1/2}).
\eeqr
Following similar arguments, we can show that
\beqr\nonumber
P_{4}&=&2n^{-1}\sum_{i=1}^n E\l[\l\{I(Y\ge Y_i)-V\r\} g'(V- V')S_0(U,U')\r.\\\nonumber
&&\l.\hspace{2cm}\cdot g(W- W')\mid Z=Z_i\r]+o_p(n^{-1/2})\\\label{P4i}
&\defby& 2n^{-1}\sum_{i=1}^n P_{4,i}+o_p(n^{-1/2}).
\eeqr
To sum up, it is shown that $c_0^{-1}\wh{\rho}(X,Y\mid Z)$ could be written as
\beqrs
&&c_0^{-1}\wh{\rho}(X,Y\mid Z)-c_0^{-1}\rho(X,Y\mid Z)\\
&=&2n^{-1}\sum_{i=1}^{n} \l\{P_{1, i}+P_{2, i}+P_{3, i}+P_{4, i} -c_0^{-1}\rho(X,Y\mid Z)\r\}+o_p(n^{-1/2}),
\eeqrs
where $P_{1,i},P_{2,i},P_{3,i}$ and $P_{4,i}$ are defined in (\ref{P1i})-(\ref{P4i}), respectively.
Thus the asymptotic normality follows.

Under the local alternative, we have $U=F(X\mid Y,Z) +n^{-1/2} \ell (X,Y,Z),$ and it is easy to verify that $\breve U \defby F(X\mid Y,Z)$, $V$ and $W$ are mutually independent.
With Taylor's expansion, we have
\beqrs
&& S_0(\wh U_i,\wh U_j)  \\
&=& \l\{ g'(\breve U_i-\breve U_j)+e^{\breve U_i-1} -e^{-\breve U_i} \r\}\Delta\breve U_i  -\l\{ g'(\breve U_i-\breve U_j)-e^{\breve U_j-1} +e^{-\breve U_j} \r\}\Delta\breve U_j \\
&+& 2^{-1}\l\{g''(\breve U_i-\breve U_j)(\Delta\breve U_i-\Delta\breve U_j)^2 +\l(e^{-\breve U_i}+e^{\breve U_i-1}\r) (\Delta\breve U_i)^2 +\l(e^{-\breve U_j}+e^{\breve U_j-1}\r) (\Delta\breve U_j)^2\r\} \\
&+& 6^{-1}\l\{g'''(\breve U_i-\breve U_j)(\Delta\breve U_i-\Delta\breve U_j)^3 +\l(e^{\breve U_i-1}-e^{-\breve U_i}\r) (\Delta\breve U_i)^3 +\l(e^{\breve U_j-1}-e^{-\breve U_j}\r) (\Delta\breve U_j)^3\r\} \\
&+&S_0(\breve U_i,\breve U_j) +o_p(n^{-1})\\
&\defby&  S_1(\breve U_i,\breve U_j)+ S_2(\breve U_i,\breve U_j)+ S_3(\breve U_i,\breve U_j)+S_0(\breve U_i,\breve U_j)+o_p(n^{-1}),
\eeqrs
where $\Delta\breve U_i = \wh U_i -\breve U_i$. Then we can write $c_0^{-1}\wh{\rho}(X,Y\mid Z)$ as
\beqrs
&&c_0^{-1}\wh{\rho}(X,Y\mid Z) \\
&=& 2^{-1} n^{-2}\sum_{i,j}S_0(\breve U_i,\breve U_j) S_0\l( V_i, V_j\r) g''(W_i-W_j)(\Delta W_i-\Delta W_j)^2\\
&&+ n^{-2}\sum_{i,j}\sum_{0\le k+l\le1} S_k(\breve U_i,\breve U_j) S_l\l( V_i, V_j\r) g'(W_i-W_j)(\Delta W_i-\Delta W_j)\\
&&+ n^{-2}\sum_{i,j}\sum_{0\le k+l\le3} S_k(\breve U_i,\breve U_j) S_l\l( V_i, V_j\r)  g\l( W_i- W_j\r)+o_p(n^{-1})\\
&\defby& 2^{-1}\wt Q_1+ \wt Q_2+ \wt Q_3+o_p(n^{-1}).
\eeqrs
With the same arguments as that in deriving $Q_1=o_p(n^{-1})$ in the proof of Theorem \ref{rho_null}, we have $\wt Q_1=o_p(n^{-1})$.

Now we deal with $ \wt Q_2$. For ease of notation, we write $\ell (X_i,Y_i,Z_i)$ as $\ell_i$ in the remaining proof. By decomposing $\Delta\breve U_i $ as $\Delta\breve U_i = \Delta U_i +n^{-1/2}\ell_i$, we have
\beqrs
\wt Q_2 &=& n^{-2}\sum_{i,j} S_1(\breve U_i,\breve U_j) S_0\l( V_i, V_j\r) g'(W_i-W_j)(\Delta W_i-\Delta W_j)+o_p(n^{-1})\\
&=& 2n^{-2}\sum_{i,j}\l\{ g'(\breve U_i-\breve U_j)+e^{\breve U_i-1} -e^{-\breve U_i} \r\}\Delta\breve U_i S_{0}\l( V_i, V_j\r) g'(W_i-W_j)\Delta W_i \\
&&-2n^{-2}\sum_{i,j}\l\{ g'(\breve U_i-\breve U_j)+e^{\breve U_i-1} -e^{-\breve U_i} \r\}\Delta\breve U_i\\
&&\hspace{2cm}\cdot S_{0}\l( V_i, V_j\r) g'(W_i-W_j)\Delta W_j+o_p(n^{-1})\\
&\defby&2\wt Q_{2,1}-2\wt Q_{2,2}+o_p(n^{-1}).
\eeqrs
$\wt Q_{2,1}$ is clearly of order $o_p(n^{-1})$ because for each fixed $i$,
\beqrs
E\l[\l.\l\{ g'(\breve U_i-\breve U_j)+e^{\breve U_i-1} -e^{-\breve U_i} \r\} S_{0}\l( V_i, V_j\r) g'(W_i-W_j)\ \r|\ (\breve U_i,V_i,W_i)\r]=0.
\eeqrs
$\wt Q_{2,2}$ is also of order $o_p(n^{-1})$ because for each $i$,
\beqrs
n^{-2}\sum_{j,k}\bigg[\l\{ g'(\breve U_i-\breve U_j)+e^{\breve U_i-1} -e^{-\breve U_i} \r\} S_{0}\l( V_i, V_j\r) g'(W_i-W_j)\l\{\indic(W_k\le W_j)-W_j\r\}\bigg]
\eeqrs
is degenerate.

Then we deal with the last quantity, $\wt Q_3$, where
\beqrs
\wt Q_3
&=&  n^{-2}\sum_{i,j} S_0(\breve U_i,\breve U_j) S_0\l( V_i, V_j\r)  g\l( W_i- W_j\r)\\
&&+n^{-2}\sum_{i,j} \sum_{k+l=1} S_k(\breve U_i,\breve U_j) S_l\l( V_i, V_j\r)  g\l( W_i- W_j\r)\\
&&+n^{-2}\sum_{i,j}\sum_{k+l=2} S_k(\breve U_i,\breve U_j) S_l\l( V_i, V_j\r)  g\l( W_i- W_j\r)\\
&&+n^{-2}\sum_{i,j}\sum_{k+l=3} S_k(\breve U_i,\breve U_j) S_l\l( V_i, V_j\r)  g\l( W_i- W_j\r)+o_p(n^{-1})\\
&\defby&\wt Q_{3,0}+\wt Q_{3,1}+\wt Q_{3,2}+\wt Q_{3,3}+o_p(n^{-1}).
\eeqrs
We simplify $\wt Q_{3,1}$ first. According to the proof of Theorem \ref{rho_null}, we have
\beqrs
\wt Q_{3,1}&=&  n^{-2}\sum_{i,j} S_1(\breve U_i,\breve U_j) S_0\l( V_i, V_j\r)  g\l( W_i- W_j\r)+o_p(n^{-1})\\
&=&2n^{-5/2}\sum_{i,j}\l\{ g'(\breve U_i-\breve U_j)+e^{\breve U_i-1} -e^{-\breve U_i} \r\}\ell_i S_0\l( V_i, V_j\r)  g\l( W_i- W_j\r)\\
&&+2n^{-3}\sum_{i,j,k} \bigg[ \l\{ g'(\breve U_i-\breve U_j)+e^{\breve U_i-1} -e^{-\breve U_i} \r\} S_0\l( V_i, V_j\r)  g\l( W_i- W_j\r)\\
&&\hspace{2cm}\cdot\l\{\frac{K_h(Z_k-Z_i)\indic(X_k\le X_i)}{ f(Z_i)}-\frac{U_i K_h(Z_k-Z_i)}{f(Z_i)}\r\}\bigg]+o_p(n^{-1})\\
&\defby&\wt Q_{3,1,1}+2\wt Q_{3,1,2}+o_p(n^{-1}).
\eeqrs
As we can see, $\frac{K_h(Z_k-Z_i)\indic(X_k\le X_i)}{ f(Z_i)}-\frac{U_i K_h(Z_k-Z_i)}{f(Z_i)}$ is of order $h^m$. Then we can derive that
\beqrs
\wt Q_{3,1,2} &=& n^{-1}\sum_{j=1}^n E \bigg[ \l\{ g'(\breve U_i-\breve U_j)+e^{\breve U_i-1} -e^{-\breve U_i} \r\} S_0\l( V_i, V_j\r)  g\l( W_i- W_j\r)\\
&&\cdot\l\{\frac{K_h(Z_k-Z_i)\indic(X_k\le X_i)}{ f(Z_i)}-\frac{U_i K_h(Z_k-Z_i)}{f(Z_i)}\r\}\ \bigg|\ (X_j,Y_j,Z_j)\bigg]+O_p(n^{-1}h^m).
\eeqrs
It can be verified that
\beqrs
&&E\l\{\frac{K_h(Z_k-Z_i)\indic(X_k\le X_i)}{ f(Z_i)}-\frac{U_i K_h(Z_k-Z_i)}{f(Z_i)}\ \bigg|\ (X_i,Z_i)\r\}\\
&=&E\l\{\frac{K_h(Z_k-Z_i)F(X_i\mid Z_k)}{ f(Z_i)}-\frac{U_i K_h(Z_k-Z_i)}{f(Z_i)}\ \bigg|\ (X_i,Z_i)\r\}\\
&=&f^{-1}(Z_i)\int K(u)\l\{F(X_i\mid Y_i,Z_i+uh)+n^{-1/2}\ell(X_i,Y_i,Z_i+uh)\r\}f(uh+Z_i)du\\
&&- f^{-1}(Z_i)\breve U_i E \l\{K_h(Z_k-Z_i)\mid Z_i\r\}-n^{-1/2}f^{-1}(Z_i)\ell_i E \l\{K_h(Z_k-Z_i)\mid Z_i\r\}.
\eeqrs
And
\beqrs
f^{-1}(Z_i)\int K(u)F(X_i\mid Y_i,Z_i+uh)f(uh+Z_i)du- f^{-1}(Z_i)\breve U_i E \l\{K_h(Z_k-Z_i)\mid Z_i\r\}
\eeqrs
is of order $h^m$ and is only a function of $(\breve U_i, Z_i)$, which is independent of $V_i$. Substituting this into $\wt Q_{3,1,2}$, we have
\beqrs
\wt Q_{3,1,2} &=& n^{-3/2}\sum_{j=1}^n E \bigg( \l\{ g'(\breve U_i-\breve U_j)+e^{\breve U_i-1} -e^{-\breve U_i} \r\} S_0\l( V_i, V_j\r)  g\l( W_i- W_j\r)\\
&&\hspace{-1cm}\cdot\l[\frac{\int K(u)\l\{\ell(X_i,Y_i,Z_i+uh)-\ell_i\r\}f(Z_i+uh)du}{ f(Z_i)}\r]\ \bigg|\ (X_j,Y_j,Z_j)\bigg)+O_p(n^{-1}h^m).
\eeqrs
Then  $\wt Q_{3,1,2}$ is clearly of order $o_p(n^{-1})$ by noting that the conditional expectation of the above display is of order $h^m$ while the unconditional expectation is zero.

Next, we deal with $\wt Q_{3,2}$. It is straightforward that
\beqrs
\wt Q_{3,2}&=&n^{-2}\sum_{i,j} S_1(\breve U_i,\breve U_j) S_1\l( V_i, V_j\r)  g\l( W_i- W_j\r)\\
&&+n^{-2}\sum_{i,j} S_2(\breve U_i,\breve U_j) S_0\l( V_i, V_j\r)  g\l( W_i- W_j\r)+o_p(n^{-1})\\
&\defby&\wt Q_{3,2,1}+\wt Q_{3,2,2}+o_p(n^{-1}).
\eeqrs
Similar to dealing with $\wt Q_{3,1,2}$, we can show that
\beqrs
\wt Q_{3,2,1}&=&2n^{-5/2}\sum_{i,j}\l\{ g'(\breve U_i-\breve U_j)+e^{\breve U_i-1} -e^{-\breve U_i} \r\}\ell_i S_1\l( V_i, V_j\r)  g\l( W_i- W_j\r)+o_p(n^{-1}).
\eeqrs
Then $\wt Q_{3,2,1}$ is of order $o_p(n^{-1})$ because $S_1\l( V_i, V_j\r) =o_p(1)$ and
\beqrs
\l\{ g'(\breve U_i-\breve U_j)+e^{\breve U_i-1} -e^{-\breve U_i} \r\}\ell_i S_1\l( V_i, V_j\r)  g\l( W_i- W_j\r)
\eeqrs
is also $o_p(1)$ with the expectation being zero.
Similar as before, we can show that
\beqrs
\wt Q_{3,2,2}&=&2^{-1}n^{-3}\sum_{i,j}\bigg[ \bigg\{g''(\breve U_i-\breve U_j)(\ell_i-\ell_j)^2 +\l(e^{-\breve U_i}+e^{\breve U_i-1}\r) \ell_i^2 \\
&&\hspace{2cm}+\l(e^{-\breve U_j}+e^{\breve U_j-1}\r) \ell_j^2\bigg\} S_0\l( V_i, V_j\r)  g\l( W_i- W_j\r)\bigg]+o_p(n^{-1})\\
&=&2^{-1}n^{-1} E\bigg[ \l\{g''(\breve U_1-\breve U_2)(\ell_1-\ell_2)^2 +\l(e^{-\breve U_1}+e^{\breve U_1-1}\r) \ell_1^2 +\l(e^{-\breve U_2}+e^{\breve U_2-1}\r) \ell_2^2\r\}\\
&&\hspace{2cm}\cdot S_0\l( V_1, V_2\r)  g\l( W_1- W_2\r)\bigg]+o_p(n^{-1})\\
&=&-n^{-1} E \l\{g''(\breve U_1-\breve U_2)\ell_1\ell_2  S_0\l( V_1, V_2\r)  g\l( W_1- W_2\r)\r\}.
\eeqrs

Now we show that $\wt Q_{3,3}=o_p(n^{-1})$. Because $(\Delta U_i)^2=o_p(n^{-1/2})$, we have
\beqrs
\wt Q_{3,3} &=&2n^{-2}\sum_{i,j} \l\{ g'(\breve U_i-\breve U_j)+e^{\breve U_i-1} -e^{-\breve U_i} \r\}\Delta U_i  S_2\l( V_i, V_j\r)  g\l( W_i- W_j\r)\\
&&+2^{-1}n^{-2}\sum_{i,j} \bigg\{g''(\breve U_i-\breve U_j)(\Delta U_i-\Delta U_j)^2 +\l(e^{-\breve U_i}+e^{\breve U_i-1}\r) (\Delta U_i)^2 \\
&&\hspace{2.5cm}+\l(e^{-\breve U_j}+e^{\breve U_j-1}\r) (\Delta U_j)^2\bigg\} S_1\l( V_i, V_j\r)  g\l( W_i- W_j\r)\\
&&+6^{-1}n^{-2}\sum_{i,j}\bigg\{g'''(\breve U_i-\breve U_j)(\Delta U_i-\Delta U_j)^3 +\l(e^{\breve U_i-1}-e^{-\breve U_i}\r) (\Delta U_i)^3 \\
&&\hspace{2.5cm}+\l(e^{\breve U_j-1}-e^{-\breve U_j}\r) (\Delta U_j)^3\bigg\}S_0\l( V_i, V_j\r)  g\l( W_i- W_j\r)+o_p(n^{-1/2}).
\eeqrs
Similar to dealing with $\wt Q_{3,1,2}$, we can obtain that $\wt Q_{3,3}=o_p(n^{-1})$.

Combining these results together, we have
\beqrs
&&c_0^{-1}\wh{\rho}(X,Y\mid Z) \\
&=& n^{-2}\sum_{i,j} S_0(\breve U_i,\breve U_j) S_0\l( V_i, V_j\r)  g\l( W_i- W_j\r)\\
&&+2n^{-5/2}\sum_{i,j}\l\{ g'(\breve U_i-\breve U_j)+e^{\breve U_i-1} -e^{-\breve U_i} \r\}\ell_i S_0\l( V_i, V_j\r)  g\l( W_i- W_j\r)\\
&& -n^{-1} E \l\{g''(\breve U_1-\breve U_2)\ell_1\ell_2  S_0\l( V_1, V_2\r)  g\l( W_1- W_2\r)\r\}  +o_p(n^{-1}).
\eeqrs
Then we can verify that $c_0^{-1}\wh{\rho}(X,Y\mid Z)$ can be written as
\beqrs
c_0^{-1}\wh{\rho}(X,Y\mid Z)
&=& \iiint\bigg\|n^{-1}\sum_{j=1}^n\bigg[ \left\{e^{it_1\breve U_j}-\varphi_{\breve U}(t_1)\right\}\left\{e^{it_2 V_j}-\varphi_{V}(t_2)\right\}e^{it_3 W_j}\\
&&\hspace{-1cm}+it_1n^{-1/2}\ell_je^{it_1\breve U_j}\left\{e^{it_2 V_j}-\varphi_{V}(s)\right\}e^{it_3 W_j}\bigg]\bigg\|^{2} \omega(t_1, t_2, t_3) d t_1 d t_2 d t_3+o_p(n^{-1}).
\eeqrs
It is clear that the empirical process
\beqrs
n^{-1/2}\sum_{j=1}^n\bigg[ \left\{e^{it_1\breve U_j}-\varphi_{\breve U}(t_1)\right\}\left\{e^{it_2 V_j}-\varphi_{V}(t_2)\right\}e^{it_3 W_j}+it_1\ell_je^{it_1\breve U_j}\left\{e^{it_2 V_j}-\varphi_{V}(t_2)\right\}e^{it_3 W_j}\bigg]
\eeqrs
converges in distribution to a complex valued gaussian process $\zeta(t_1,t_2,t_3)$ with mean function
\beqrs
E\left[it_1\ell(X,Y,Z)e^{it_1\breve U}\left\{e^{it_2 V}-\varphi_{V}(t_2)\right\}e^{it_3 W}\right],
\eeqrs
and covariance function $\cov\{\zeta(t_1,t_2,t_3),\overline{\zeta(t_{10},t_{20},t_{30})}\}$ given by
\beqr\label{cov}
\l\{\varphi_{ U}(t_1-t_{10})-\varphi_{ U}(t_1)\varphi_{ U}(-t_{10})\r\}\l\{\varphi_{V}(t_2-t_{20})-\varphi_{V}(t_2)\varphi_{V}(-t_{20})\r\}\varphi_{W}(t_3-t_{30}).
\eeqr
Therefore, by employing empirical process technology, we can derive that
\beqrs
c_0^{-1}n\wh{\rho}(X,Y\mid Z) \overset{d}{\longrightarrow} \iiint\left\|\zeta(t_1,t_2,t_3)\right\|^{2} \omega(t_1,t_2,t_3) d t_1 d t_2 d t_3.
\eeqrs
Hence we conclude the proof for local alternatives.


\hfill$\fbox{}$

\subsection{Proof of Theorem \ref{rho_multivariate}}
Firstly, $\x\hDash \y\mid \z$ is equivalent to $(X_1,\ldots,X_p)\hDash \y\mid \z$. According to Proposition 4.6 of \cite{cook2009regression}, it is also equivalent to
\beqrs
X_1\hDash \y\mid \z,\quad X_2\hDash \y\mid (\z,X_1),\quad \ldots,\quad X_p\hDash \y\mid (\z,X_1,\ldots,X_{p-1}).
\eeqrs
Following similar arguments for proving the equivalence between $X\hDash Y\mid Z$ and $U\hDash V\mid Z$ in the proof of Proposition \ref{equivalence},
the above conditional independence series are equivalent to
\beqrs
\wt U_1\hDash \y\mid\z,\quad \wt U_2\hDash \y\mid (\z,X_1),\quad \ldots,\quad \wt U_p\hDash\y\mid (\z,X_1,\ldots,X_{p-1}).
\eeqrs
According to the proof of Proposition \ref{equivalence}, we know that $\wt U_k\hDash\y\mid (\z,X_1,\ldots,X_{k-1})$  is equivalent to
$\wt U_k\hDash\y\mid  (\z,\wt U_1,\ldots,\wt U_{k-1})$ for $k=1,\ldots,p-1$. Hence the conditional independence series hold if and only if
\beqrs
\wt U_1\hDash \y\mid \z,\quad \wt U_2\hDash \y\mid (\z,\wt U_1),\quad \ldots,\quad \wt U_p\hDash \y\mid (\z,\wt U_1,\ldots,\wt U_{p-1}).
\eeqrs
Then by applying Proposition 4.6 of \cite{cook2009regression} again, we know that $(X_1,\ldots,X_p)\hDash \y\mid \z$ is equivalent to $\u\hDash \y\mid \z$. Furthermore, with the same arguments for dealing with $\y$, we can obtain that it is additionally equivalent to $\u\hDash \v\mid \z$.
Besides, with the fact that $\u \hDash\z$ and $\v\hDash\z$, we can get the conditional independence $\x\hDash\y\mid\z$ is equivalent to the mutual independence of $\u$, $\v$ and $\z$.
Therefore, the proof is completed by following similar arguments with the proof of Proposition \ref{equivalence}.
\hfill$\fbox{}$

\subsection{Proof of Theorem \ref{T_n_null}}

Following the proof of Theorem \ref{rho_null},
we denote by $\wt g(\u_1,\u_2)=e^{-\| \u_1- \u_2\|_1}$, $\wt g(\v_1,\v_2)=e^{-\| \v_1- \v_2\|_1}$
and $\wt g(\w_1,\w_2)=e^{-\| \w_1- \w_2\|_1}$.  Then we have
\beqrs
S_{\u}(\u_1,\u_2) &=& E\l\{\wt g(\u_1,\u_2)+\wt g(\u_3,\u_4)-\wt  g(\u_1,\u_3)-\wt g(\u_2,\u_3)\mid (\u_1,\u_2)\r\}, \\
S_{\v}(\v_1,\v_2) &=& E\l\{\wt g(\v_1,\v_2)+\wt g(\v_3,\v_4)- \wt g(\v_1,\v_3)-\wt g(\v_2,\v_3)\mid (\v_1,\v_2)\r\}.
\eeqrs
Therefore, $\wh \rho(\x, \y \mid\z)$ can be written as
\beqrs
\wh \rho(\x, \y \mid\z)= n^{-2}\sum_{i,j} \l\{S_{\u}(\wh \u_i,\wh \u_j) S_{\v}(\wh \v_i,\wh \v_j) \wt g(\wh \w_i,\wh \w_j)\r\}.
\eeqrs
With Taylor's expansion, when $nh^{4m}\to 0$,
$nh^{2(r+p-1)}/\log^2(n)\to \infty$, under conditions $2'$ and $3'$, we have
\beqrs
\wt g(\wh \u_1,\wh \u_2) = \wt g(\u_1,\u_2)+\sum_{k=1}^3(\Delta\u_1\trans,\Delta\u_2\trans)^{\otimes k}D^{\otimes k}\wt g(\u_1,\u_2)+o_p(n^{-1}),
\eeqrs
where $\mathbf A^{\otimes k}$ denotes the $k$-th Kronecker power of the matrix $\mathbf A$,
$\Delta \u_i = \wh \u_i -\u_i$ and
\beqrs
D^{\otimes k}\wt g(\u_1,\u_2) = \frac{\partial ^k \wt g(\u_1,\u_2)}{\{\partial (\u_1\trans,\u_2\trans)\trans\}^{\otimes k}}.
\eeqrs
In addition, we can expand $\wt g(\wh \u_1,\u_2)$ as
\beqrs
\wt g(\wh \u_1,\u_2) = \wt g(\u_1,\u_2)+\sum_{k=1}^3(k!)^{-1}(\Delta\u_1\trans,{\bf 0}\trans)^{\otimes k}D^{\otimes k}\wt g(\u_1,\u_2)+o_p(n^{-1}).
\eeqrs
Therefore, by the definition of $S_{\u}(\u_1,\u_2) $, we have
\beqrs
&& S_{\u}(\wh\u_i,\wh\u_j) \\
&=&E\l\{\wt g(\u_i,\u_j)+\wt g(\u,\u')-\wt  g(\u_i,\u)-\wt g(\u,\u_j)\mid (\u_i,\u_j)\r\} \\
&+& \sum_{k=1}^3(k!)^{-1}E\Big\{(\Delta\u_i\trans,\Delta\u_j\trans)^{\otimes k}D^{\otimes k}\wt g(\u_i,\u_j)
-(\Delta\u_i\trans,{\bf 0}\trans)^{\otimes k}D^{\otimes k}\wt g(\u_i,\u)\\
&&\hspace{1cm}-({\bf 0}\trans,\Delta\u_j\trans)^{\otimes k}D^{\otimes k}\wt g(\u,\u_j)\mid \u_i,\u_j\Big\}+o_p(n^{-1}),\\
&\defby& \sum_{k=0}^3\wt S_k(\u_i,\u_j)+o_p(n^{-1}),
\eeqrs
where $\wt S_0(\u_i,\u_j)=S_{\u}(\u_i,\u_j)$ and $\wt S_k(\u_i,\u_j), k=1,2,3$ are defined obviously.
Similarly, when
$nh^{2(r+q-1)}/\log^2(n)\to \infty$, and $nh^{4m}\to 0$,
we can expand $S_{\v}(\wh\v_i,\wh\v_j) $ as
\beqrs
S_{\v}(\wh\v_i,\wh\v_j)
&\defby&\sum_{k=0}^3\wt S_k(\v_i,\v_j)+o_p(n^{-1}),
\eeqrs
and it follows that
\beqrs
\wh \rho(\x, \y \mid\z)&=& n^{-2}\sum_{i,j} \sum_{0\le k+l\le3}\wt S_k(\u_i,\u_j)\wt S_l(\v_i,\v_j)\wt g( \w_i, \w_j) \\
&&+ n^{-2}\sum_{i,j} \sum_{0\le k+l\le2}\wt S_k(\u_i,\u_j)\wt S_l(\v_i,\v_j)(\Delta\w_i\trans,\Delta\w_j\trans) D \wt g(\w_i,\w_j) \\
&&+ 2^{-1}n^{-2}\sum_{i,j} \sum_{0\le k+l\le1}\wt S_k(\u_i,\u_j)\wt S_l(\v_i,\v_j)(\Delta\w_i\trans,\Delta\w_j\trans)^{\otimes 2}D^{\otimes 2}\wt g(\w_i,\w_j)\\
&&+ 6^{-1}n^{-2}\sum_{i,j}  \wt S_0(\u_i,\u_j)\wt S_0(\v_i,\v_j)(\Delta\w_i\trans,\Delta\w_j\trans)^{\otimes 3}D^{\otimes 3}\wt g(\w_i,\w_j)+o_p(n^{-1})\\
&\defby& Q_1'+ Q_2'+ Q_3'+ Q_4'+o_p(n^{-1}).
\eeqrs
Then following similar arguments in the proof of Theorem \ref{rho_null},
we have
$Q_2'$, $Q_3'$ and $Q_4'$ are all of order $o_p(n^{-1})$ and $Q_1'$ equals
$n^{-2}\sum_{i,j} S_\u(\u_i,\u_j) S_\v(\v_i,\v_j)\wt g( \w_i, \w_j)+o_p(n^{-1})$.
Combing these results, we have
\beqrs
\wh \rho(\x, \y \mid\z)=n^{-2}\sum_{i,j} S_\u(\u_i,\u_j) S_\v(\v_i,\v_j)\wt g( \w_i, \w_j)+o_p(n^{-1}),
\eeqrs
where the right hand side is a first order degenerate V statistics. Thus by applying Theorem 6.4.1.B of \cite{serfling2009approximation},
\beqrs
n\wh \rho(\x, \y \mid\z)\xrightarrow{d}  \sum_{j=1}^{\infty}\lambda_j\chi_{j}^{2}(1)
\eeqrs
where $\chi_{j}^2(1)$, $j=1, 2, \dots$ are independent $\chi^2(1)$ random variables, and $\lambda_j$, $j=1, 2, \dots$ are the  corresponding eigenvalues of $\wt h(\u, \v, \w; \u', \v', \w')$. Therefore, the proof is completed.
\hfill$\fbox{}$

\subsection{Proof of Theorem \ref{rho_discrete}}
It suffices to show that $X\hDash Y\mid Z$ if and only if $U\hDash V\mid Z$ because $U\hDash V\mid Z$ is equivalent to $U,V$ and $Z$ are mutually independent under $U\hDash Z$ and $V\hDash Z$.

We only show that $X\hDash Y\mid Z$ if and only if $U\hDash Y\mid Z$, because similar arguments will yield that it is also equivalent to $U\hDash V\mid Z$.
It is quite straightforward that $X\hDash Y\mid Z$ implies $U\hDash Y\mid Z$. While when $U\hDash Y\mid Z$, we have
for each $Z=z$, $u$ and $y$ in the corresponding support,
\beqrs
\pr(U\le u,Y\le y\mid Z=z) = uF_{Y\mid Z}(y\mid z).
\eeqrs
Substituting $U=(1-U_X) F_{X\mid Z}(X-\mid Z)+U_X F_{X\mid Z}(X\mid Z)$ into the above equation, with some straight calculation, the left hand side is
\beqrs
& \pr\l\{\pr(\wt X<X\mid X,\wt Z = z)+\pr(\wt X=X\mid X,\wt Z = z)U_X\le u,Y\le y\mid Z=z\r\}\\
=& \pr\l\{\pr(\wt X<X\mid X,\wt Z = z)+\pr(\wt X=X\mid X,\wt Z = z)U_X\le u,F_{Y\mid X,Z}(y\mid X,z)\mid Z=z\r\}.
\eeqrs
Because $U_X$ is standard uniformly distributed, we obtain
\beqrs
E\l[ g\l\{\frac{u-\pr(\wt X< X\mid X,\wt Z=Z=z)}{\pr(\wt X= X\mid X,\wt Z=Z=z)}\r\}F_{Y\mid X,Z}(y\mid X,z)\r]=uF_{Y\mid Z}(y\mid z),
\eeqrs
where $g(\cdot)$ is the cumulative distribution function of a standard uniformly distributed random variable.
Now assume that conditional on $Z=z$, the support of $X$ is $\{x_1,\ldots,x_N\}$, where $x_1<\ldots<x_N$.
Therefore, when $0<u<F_{X\mid Z}(x_1\mid z)$,
the expectation in the above equation is
\beqrs
&&\sum_{i=1}^N g\left\{\frac{u-\pr\left(X<x_{i}\mid Z=z\right)}{\pr\left(X=x_{i}\mid Z=z\right)}\right\} F_{Y\mid X,Z}(y\mid x_i,z) \pr\left(X=x_{i}\mid Z=z\right)\\
&=&\frac{u-\pr\left(X<x_{1}\mid Z=z\right)}{\pr\left(X=x_{1}\mid Z=z\right)}\pr\left(X=x_{1}\mid Z=z\right)F_{Y\mid X,Z}(y\mid x_1,z)\\
&=&uF_{Y\mid X,Z}(y\mid x_1,z).
\eeqrs
The expectation equals $uF_{Y\mid Z}(y\mid z)$. That is, $F_{Y\mid X,Z}(y\mid x_1,z)=F_{Y\mid Z}(y\mid z)$.

When $F_{X\mid Z}(x_1\mid z)<u<F_{X\mid Z}(x_2\mid z)$, we can calculate the expectation as
\beqrs
&&\sum_{i=1}^N g\left\{\frac{u-\pr\left(X<x_{i}\mid Z=z\right)}{\pr\left(X=x_{i}\mid Z=z\right)}\right\} F_{Y\mid X,Z}(y\mid x_i,z) \pr\left(X=x_{i}\mid Z=z\right)\\
&=&F_{Y\mid X,Z}(y\mid x_1,z) \pr\left(X=x_{1}\mid Z=z\right)+\l\{u-\pr\left(X<x_{2}\mid Z=z\right)\r\}F_{Y\mid X,Z}(y\mid x_2,z).
\eeqrs
Since we have shown that $F_{Y\mid X,Z}(y\mid x_1,z)=F_{Y\mid Z}(y\mid z)$, with the fact that the expectation equals $uF_{Y\mid Z}(y\mid z)$, we can get $F_{Y\mid X,Z}(y\mid x_2,z)=F_{Y\mid Z}(y\mid z)$.

Similarly, we can obtain that $F_{Y\mid X,Z}(y\mid x_k,z)=F_{Y\mid Z}(y\mid z)$, $k=3,\ldots, N$. Consequently, we have $F_{Y\mid X,Z}(y\mid x,z)=F_{Y\mid Z}(y\mid z)$ for all $x,y$ and $z$ in their support. That is, $X\hDash Y\mid Z$. Therefore, the proof is completed.

\hfill$\fbox{}$

\section{Additional Simulations Results}
We consider the directed acyclic graph
with 5 nodes, i.e., $X=(X_1,\dots, X_5)$, and only allow directed edge from
$X_i$ and $X_j$ for $i<j$. Denote the adjacency matrix $A$. The existence of
the edge follows a Bernoulli distribution, and we set $\pr(A_{i,j} = 1)=0.4$, for
$i<j$. When $A_{i,j}=1$, we replace $A_{i,j}$ with independent realizations of
a uniform $U(0.1,1)$ random variable. The value of the first random variable
$X_1$ is randomly sampled from some distribution $\wt P$. Specifically,
\beqrs
\ep_1 \sim \wt P,\mbox{ and } X_1 = \ep_1.
\eeqrs
The value of the next nodes is
\beqrs
\ep_j \sim \wt P,\mbox{ and } X_j=\sum_{k=1}^{j-1}A_{j,k}X_k+ \ep_j
\eeqrs
for $j=1,\cdots, p$. All random errors $\ep_1, \dots,
\ep_p$ are independently sampled from the distribution $\wt P$. We consider two
scenarios, where $\wt P$  follows either  normal distribution $N(0,1)$  or
uniform distribution $U(0,1)$. We compare our proposed conditional independence test (denoted by ``CIT'')
with other popular conditional dependence measure. They are,
respectively, the partial correlation \citep[denoted by
``PCR''], conditional mutual information \citep[denoted by
``CMI'']{scutari2010}, and the KCI.test \citep[denoted by
``KCI'']{zhang2011kci}. We set the sample size  $n=50$, 100, 200 and  $300$. The
true positive rate  and false positive rate  for the four different
tests are reported in Tables \ref{causal1}, from which
we can see that as the sample size increases, the true positive rate of the proposed
method steadily grows, and the proposed method outperforms the other tests,
while the false positive rate remains under control with slightly decrease.

\begin{table}[h]
	\centering
	\caption{The true positive rate  and false positive rate  for the causal discovery of the directed acyclic
		graph with different tests}
	\label{causal1}
	\scalebox{.95}{\begin{tabular}{lcccccccc}
			\hline
			\multicolumn{1}{c}{Samples}       & 50      & 100      & 200     & 300     & 50      & 100      & 200      & 300     \\ \hline
			\multicolumn{1}{c}{Tests}  & \multicolumn{8}{c}{\rule{0pt}{3ex}   $\wt P \sim N(0,1)$}                                                                                              \\ \hline
			& \multicolumn{4}{c}{true positive rate} & \multicolumn{4}{c}{false positive rate} \\
			CIT      & 0.555 & 0.658 & 0.734 & 0.789        & 0.117 & 0.112 & 0.107 & 0.103   \\
			PCR            & 0.479 & 0.489 & 0.546 & 0.589     & 0.101 & 0.110 & 0.130 & 0.135       \\
			CMI & 0.472 & 0.530 & 0.604 & 0.590    & 0.097 & 0.127 & 0.143 & 0.150  \\
			KCI  & 0.360 & 0.516 & 0.592 & 0.634     & 0.072 & 0.135 & 0.144 & 0.168\\ \hline
			\multicolumn{1}{c}{Tests}  & \multicolumn{8}{c}{\rule{0pt}{3ex}   $\wt P \sim U(0,1)$}                                                                                              \\ \hline
			\multicolumn{1}{c}{}           & \multicolumn{4}{c}{true positive rate} & \multicolumn{4}{c}{false positive rate} \\
			CIT     & 0.468 & 0.587 & 0.734 & 0.736   & 0.070 & 0.099 & 0.095 & 0.113  \\
			PCR          & 0.469 & 0.526 & 0.588 & 0.545   & 0.111 & 0.129 & 0.140 & 0.140     \\
			CMI  & 0.497 & 0.568 & 0.566 & 0.633   & 0.103 & 0.127 & 0.138 & 0.149   \\
			KCI   & 0.386 & 0.458 & 0.523 & 0.564  & 0.082 & 0.099 & 0.123 & 0.122 \\ \hline
	\end{tabular}}
\end{table}

\bibliographystyle{refstyle.bst}
\bibliography{reference}

\end{document}